\documentclass[a4paper]{article}
\usepackage[dvips]{graphicx}
\usepackage{amsmath,amssymb}
\usepackage{epsf}

\newcommand{\bs}{\begin{subequations}}
\newcommand{\es}{\end{subequations}}

\numberwithin{equation}{section}

\def \myfigures #1#2#3#4#5#6#7#8
{\begin{figure}
    \begin{center}
        \includegraphics[width=#2 \textwidth]{#1.eps}
        \hfill
        \includegraphics[width=#6 \textwidth]{#5.eps}
        \parbox[t]{#4\textwidth}{\caption {#3}\label{#1}}
        \hfill
        \parbox[t]{#8\textwidth}{\caption {#7}\label{#5}}
    \end{center}
\end{figure} }

\begin{document}

\title{Numerical Modeling of Charged Black Holes\\ with Massive Dilaton
\thanks{A talk given at V International Congress on Mathematical Modeling, 
Sep. 30 -- Oct. 6, Dubna, Russia, 2002: http://www.jinr.ru/vicmm/}}
         
\author{T.L. Boyadjiev\thanks{Joint Institute for Nuclear Research, Dubna, Russia. E-mail: todorlb@jinr.ru},\ \ P.P. Fiziev\thanks{Faculty of Physics, University of Sofia, Sofia, Bulgaria. E-mail: fiziev@phys.uni-sofia.bg}}

\date{}
\maketitle


\begin{abstract}
In this paper the static, spherically symmetric and electrically charged black hole solutions in Einstein-Born-Infeld gravity with massive dilaton are investigated numerically.

The Continuous Analog of Newton Method (CANM) is used to solve the corresponding nonlinear multipoint boundary value problems (BVPs). The linearized BVPs are solved numerically by means of collocation scheme of fourth order.

A special class of solutions are the extremal ones. We show that the extremal horizons within the framework of the model satisfy some nonlinear system of algebraic equations. Depending on the charge $q$ and dilaton mass $\gamma$, the black holes can have no more than three horizons. This allows us to construct some Hermite polynomial of third order. Its real roots describe the number, the type and other characteristics of the horizons.
\end{abstract}



\section {Introduction}

Being a product of strong nonlinearity of the modern theory of gravitation, the black holes (BH) represent a significant challenge since their theoretical discovery within the framework of a general theory of relativity (GTR) \cite {chandra98}. Till now BH are not a subject of a direct experimental analysis and at present there is no indisputable example of observed BH with their critical property -- {\em the existence of an event horizon} \cite{RN}. Moreover, there exist serious physical arguments supporting Einstein-Dirac's assertion \cite{Einstein} that the horizons are not a physically admissible notion. In GTR there also exist a large number of solutions of the corresponding physical problems without horizons at all. A more detailed discussion of these problems can be found in \cite{PF}. In this complicated situation one needs a trustworthy methods of both theoretical and experimental study of BH to make a serious and well founded conclusions about the real physical meaning of the solutions with horizons and their ability to describe the physical reality.

The modern theories of fundamental interactions, such as the theories of space-time with torsion, dilaton gravity, Born-Infeld electrodynamics, supergravity, superstrings, superbrane and M-theory have multiply enriched the set of different BH solutions. These models have not yet reach their final form and do not have a status of phenomenologically justified physical theory, nevertheless, they present some theoretical interest at least as a mathematical examples of nonlinear theories.\\

From  physical point of view BH can be divided into two basic classes:
\begin {enumerate}
    \item Macroscopic BH, with the mass varying from Chandrasekhar's mass $\sim 3 M_{\odot}$ to mass $\sim 10^6 - 10^{11} M_{\odot}$, {\it i.e.}, about mass of a Galaxy;
    \item Microscopical BH, whose mass has the order of that of elementary particles of the Standard model.
\end {enumerate}

If today there are some preliminary indications that the macroscopic BH may be located in active cores of galaxies, in the (not active) core of our Galaxy, in quasars, and also as a dark components in the systems of X-ray star pairs \cite {Chrusciel}, then for the time being we have not at disposal any phenomenological indications about an existence of microscopical BH. Until recently the basic reason for that was the circumstance, that the microscopical BH represent a non-linear formation  in 4D space-time, whose size is of order of the Planck length $\sim 10^{-33}cm$. This is far outside of limits of experimental accessibility even in distant future.

During the last years the situation in that plan has varied a little: miscellaneous theoretical models in the frameworks of
superstrings and superbranes augmented essentially the anticipated
size of possible  microscopical BH at the expense of usage of (at
present not observed) higher dimensions of space-time. Therefore
now one considers the opportunities for discovery of microscopical
BH on the future accelerators like LHC, VLHC, NLC at energies
about several TeV (see, for example, \cite {LHC_ch, LHC_eg} and
the references therein), and the generation of such kind BH in the
cosmic rays also (see, for example, \cite{KL}). All these make the problem for more detailed study of BH's structure in the composite modern non-linear theories with many interacting
fields in 4D non-Euclidean space-time actual.

A new feature of BH in these theories (even in presence of only
one additional scalar field) appears to be the principled
opportunity for appearance of several horizons of miscellaneous
types with different space-time structure between them \cite {B}
--- \cite{B_b}. Nevertheless, there are no convincing arguments
that one can consider the space-time domains between the various
horizons as a real co-existing structures in a physically
meaningful way, the investigation of such solutions at present
becomes a part of the general BH problem.

BH with three horizons arise even in the simplest generalizations
of GTR -- the GTR with a cosmological constant \cite{HSN}, where
historically for the first time the necessity to extend in
appropriate way the concept of event horizon and to consider a
model of space-time with complicated causal structure was suggested. An
appearance of three horizons in space-time was obtained earlier
also in the models of space-time with torsion \cite{Hehl}.

The wide classes BH with many horizons exist in various models of
the dilaton gravity (see \cite {no_01} and numerous references
there).

BH with two event horizons were obtained numerically in the
Einstein--Maxwell theory with massless dilaton \cite {EMD, hh92}.
There was pronounced conjecture, that the presence of a massive
dilaton in such models BH with three horizons also are possible,
which was numerically demonstrated in \cite {jinr01_221} within
the Einstein-Born-Infeld model.

Is is well-known that ``usual'' BH in GTR evaporate due to quantum generation of particles by their strong gravitational field. This phenomenon was discovered by Hawking and provoked a rough progress of the quantum theory of BH. Hawking also remarked \cite{H1993}, that for ``exotic'' BH with many horizons a reverse phenomenon
with respect to vaporization can take place. A simple example is
the almost singular Narai BH \cite {Narai}. That is why BH with
many horizons represent a special concern for quantum theories and
were studied by many authors (see references in \cite {no_01}).

Common feature of all BH models with several horizons is their strong nonlinearity, which, as a rule, makes impossible their exact analytical description. As a result especially actual is the problem of developing an adequate numerical methods for obtaining the solutions with several horizons. Based on our previous studies \cite {jinr01_221, jinr02_1}, here we offer and apply such general method, which we illustrate using the specific BH example in the Einstein-Born-Infeld model with massive dilaton. This model appears to be suitable polygon for development of numerical methods, which one subsequently can apply in other theories. In the present work, which is extension of \cite{jinr01_221}, we present the formulation and the numerical algorithms of solving of the problem. It is shown that the correct formulation of a BVP for BH equations (and consequently the performance of the method of solution) is essentially determined by both the number and the type of horizons.

We solve the non-linear BVPs using iterative methods based on the continuous analog of the Newton method \cite {pazppsl_99} in combination with method of collocation for solving of arising linearized problems. Compared with the methods based on solving of a Cauchy problem (see, for example, \cite {hh92}), such approach has definite advantages.

\section {Formulation of the BVP} \label {blhol1}

The dimensionless BH equations have the form (the marks below are
similar to those used in \cite {jinr01_221, jinr02_1}): \bs \label
{eqns}
    \begin {gather}
        -f \,' + F (r, f, \varphi, \varphi \,') = 0 \,, \label {metric} \\
        -f \left(\varphi\,'' + \frac {1}{r} \varphi\,' \right) +
        \Phi (r, \varphi, \varphi\,')= 0. \label {dilaton}
    \end {gather}
\es \noindent Here $f(r)$ is the function involved in the metric
space-time
\begin {displaymath}
    ds^2 = -f (r)\,e^{2 \,\delta(r)}\, dt\,^2 + f^{-1}(r)\,d\,r^2 + r^2 d \,\Omega,
\end {displaymath}
and $\varphi (r)$ represents the dilaton field. The radial coordinate $r \in [ R_l, \infty)$, where the constant $R_l > 0$, and ``right hands'' $F$ and $\Phi$ are set through expressions
\begin {displaymath}
    F \equiv \frac {1-f} {r} + 2 \exp\{ {2 \alpha \varphi}\}
    \frac {r^2 -\sqrt {r^4 + q^2}} {r} - r \gamma^2 V (\varphi) - r\, f \varphi^{\prime 2},
\end {displaymath}
    \begin {align*}
        \Phi \equiv \left [r \gamma^2 V (\varphi) -\frac {1} {r} - 2\exp {\{2\alpha \varphi
        \}} \frac {{r^2 - \sqrt {r^4 + q^2}}} {r} \right] \varphi^{\prime} +
        \frac {\gamma^2} {2} V^{\prime} (\varphi) \\ -
        2 \alpha \exp {\{2\alpha \varphi \}} \frac {{r^2 - \sqrt {r^4 + q^2}}} {{r^2}}.
    \end {align*}
\noindent In these expressions the parameter $q$ corresponds to the electric charge of BH, and $\gamma^2$ is a dilaton mass, $V (\varphi)$ --  potential of the dilaton field. The choice of the sign of coupling constant $ \alpha = \pm 1$ determines the sign of dilaton field: $\alpha = -1$  corresponds to $ \varphi(r) > 0$, whereas $\alpha = 1$ corresponds to $\varphi (r) < 0$. The case $\gamma=0$ corresponds to BH with massless dilaton, investigated in works \cite {tt00, tt01}. Let us remark that in \cite {jinr01_221} the equations \eqref{eqns} are written in a different way by introducing an additional variable $m(r) = r\, \left(1 - f(r) \right)/2$ (local mass).

For some given solution $\varphi(r)$ of problem \eqref{eqns} it is necessary to solve the following Cauchy problem
\begin {equation} \label {delta}
    \delta\,' + r \,\varphi\,'\,^2 = 0, \quad \lim \limits_{r \to \infty} \delta (r) =0\,,
\end {equation}
for the metrical function $\delta (r)$.

As usual (see, for example, \cite {chandra98}), positive zeroes $R_{h, n}$, $n = 1,2 ..., N_h \ge 1$ ($N_h$ is a number of horizons) of metric function $f (r)$ we shall call BH event horizon. We will show below that in the BH model under consideration, depending on values of electric charge $q$ and mass $\gamma$, no more than $N_h=3$ horizons may exist. Greatest of them $R_h$ we shall call an exterior (physical) horizon, and extreme left $R_l \leq R_h$ --- an internal horizon.

For the BVP related to equations \eqref{eqns} to be closed, it is necessary to formulate respective boundary conditions.

First of all we shall note that for $ \gamma \ne 0$ on the right end when $r\to \infty$ the asymptotic conditions take place:
\begin {itemize}
    \item[$\surd$] for metric function:
        \begin {equation} \label {bcr1}
            f (r) \to 1-\frac{2M_{\infty}}{r}+\frac{q^2}{r^2}\,;
        \end {equation}
    \item[$\surd$] for dilaton field:
        \begin {equation} \label {bcr2}
            \varphi (r) \to -\frac {\alpha q^2} {\gamma^2 r^4}.
        \end {equation}
\end {itemize}
\noindent In formula \eqref{bcr1} $M_{\infty}$ indicates the BH Keplerian-like mass.

The posing of boundary conditions on the right end, and hence the appropriate numerical method for the respective BVP essentially depend on both the kind and the number of horizons. Let us consider in detail the basic cases of possible boundary conditions.

\subsection {BH with regular event horizon} \label {reg_bh}

We shall refer to the horizon $R_h$ as a regular one, if the derivative
\begin{equation} \label{feq0}
    f^{\prime} (R_h) \ne 0\,.
\end{equation}
Let us consider the formulation of BVP for BH with single regular horizon. For $N_h=1$ the semi-axis $(0, \infty)$ is splitted by the point $r = R_h$ into two areas: internal $D_{int} \equiv (0, R_h)$ and external $D_{ext} \equiv (R_h, \infty)$. In the area $D_{ext}$ BVP for equations \eqref{eqns} is solved using the ``standard'' condition on the horizon
\begin {equation} \label {hor1}
    f (R_h) = 0\,.
\end {equation}
Apparently the point $R_h$ is a point of degeneration of the Eq.\ \eqref{dilaton}. To ensure the regularity of solutions on the horizon, it is necessary the equality 
\begin {displaymath}
    \Phi (R_h, \varphi_h, \varphi^{\prime}_ h) = 0\,,
\end {displaymath}
to be fulfilled, which has the following  detailed form
\begin {multline} \label {reg1}
    \left [\gamma^2 V (\varphi_h) - \frac {1} {R_h} - 2 \ \exp {\{2\alpha \varphi_h
    \}} \frac {R_h^2 - \sqrt {R_h^4 + q^2}} {R_h} \right] \varphi^{\prime}_h - \\
    \frac {\gamma^2} {2} V^{\prime}(\varphi_h) - 2 \ \alpha \exp {\{2\alpha \varphi_h
    \}}\frac {R_h^2 - \sqrt {R_h^4 + q^2}} {R_h^2} = 0\,.
\end {multline}
\noindent Here and henceforth the subscript $h$ means that the value of the respective function is calculated at the point $R_h$.

\begin {figure} 
    \begin {center}
        \includegraphics[totalheight=5.4cm,keepaspectratio]{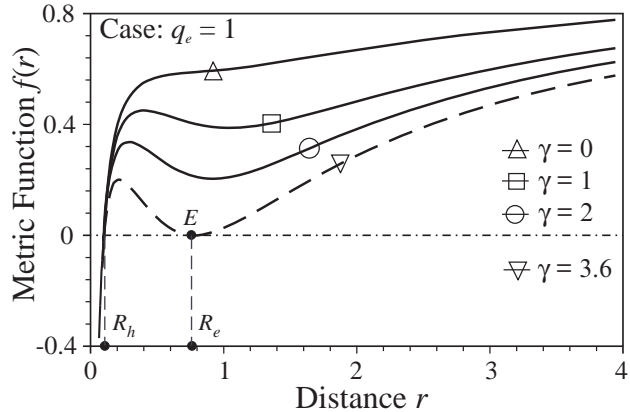} \caption {Metric function $f (r)$ for several values of dilaton mass $\gamma$}\label {fig1} 
    \end{center}
\end {figure}

Let us point out that for a given mass $M_{\infty}$ the boundary problem \eqref{eqns} -- \eqref{reg1} is a problem with free boundary \cite{vab_87}, because the point $R_h$ is {\it a priori} unknown. In many cases, however, (see discussions in \ref{discussion}) from the computational point of view, it is more convenient to set the magnitude $R_h > 0$ and solve the problem in $D_{ext}$ with a fixed boundary. Thus, mass $M_\infty$ is determined from the asymptotic \eqref{bcr1}.

The numerical method of solution of BVP for BH with given event horizon is presented in Section \ref{fixed}.

Visual examples of metric function $f(r)$ for $R_h = 0.1$ and given values of dilaton mass $\gamma=0$ (curve noted by $\triangle$), $\gamma=1$ (curve $\square$), and $\gamma = 2$ (curve $ \lozenge$) are introduced in Fig.\ \ref{fig1}. Solution $\triangle$ corresponds to BH with a massless dilaton considered in articles \cite {tt00, tt01}.

Let us note that in a series of works (see, for example, \cite {tt00, tt01}) to formulate the BVP for BH the authors confined themselves within the scope of \eqref{feq0}.

Let us consider the formulation of BVP in the internal area $ D_{int} \equiv (0, R_h)$. Let $(f_{+}(r), \varphi_{+}(r), R_h)$ be the solution of problem \eqref{eqns} --- \eqref{reg1} for $r \ge R_h$ and some given $q$ and $\gamma$. Let us suppose that Eqs. \eqref{eqns} are fulfilled as well at $r < R_h$. To obtain the solution $f_{-}(r)$, $\varphi_{-}(r)$, $R_l$
in some subdomain $r\in (R_l, R_h)$ depending on argument $R_l > 0$ we require a continuity of functions $f(r)$ and $\varphi(r)$ at the point $R_h$. It leads to
\begin {eqnarray}
    f_{-} (R_h) &=& 0\,, \label {bcl1} \\
    \varphi_{-} (R_h) &=& \varphi_{+} (R_h)\,, \label {bcl2} \\
    \varphi\,'_{-} (R_h) &=& \varphi\,'_{+} (R_h)\,. \label {bcl3}
\end {eqnarray}

For a closed boundary problem in $(R_l, R_h)$ one more boundary condition is necessary. Having in mind the absence of horizons for $0 < r < R_h$ the choice of this condition is arbitrary enough. For example, such condition can be
\begin {equation} \label {bcl4}
    |f_{-} (R_l) | = 1\,.
\end {equation}
\noindent Since parameter $ R_l$ is {\it a'priori} unknown, problem \eqref{eqns}, \eqref{bcl1} -- \eqref{bcl4} appears to be a problem with free left boundary. Numerical algorithm for
solving such BVPs is presented below in Section \ref{freebound}.

\subsection {BH with extremal horizons} \label {extrhor}

We shall say that BH has at point $R_e $ an extremal  event horizon if the conditions
\begin {equation} \label {extr1}
        f (R_e) = 0\,,
\end {equation}
$$ f\,'(R_e) = 0\,,$$
are satisfied.

An example of solution $ f(r)$ with extreme exterior horizon $R_h \approx 0.776 $ and internal $ R_l = 0.1$ for $q=1$ and $ \gamma=3.6 $ is shown on Fig.\ \ref{fig1} (curve noted by $\triangledown $).

Using Eq.\ \eqref{metric} we can write the last expression in explicit form
\begin {equation}
    1 + 2 \exp { \{2\alpha \varphi_e \} }(R_e^2 -\sqrt {R_e^4 + q^2}) -
    R_e^2 \gamma^2 V (\varphi_e) = 0\,, \label {extr2}
\end {equation}
\noindent where subscript $e$ means that the respective magnitude concerns the extreme horizon.

In order to be regular on the extreme horizon the solution has also to satisfy a constrain of kind \eqref{reg1}. In this case it simply becomes
\begin {equation}
    \frac {\gamma^2} {2} V \, ' (\varphi_e) - 2 \alpha \exp {\{2\alpha \varphi_e
    \}} \frac {R_e^2 - \sqrt {R_e^4 + q^2}} {R_e^2} = 0.
    \label {rege}
\end {equation}

For given values of physical quantities $q$ and $\gamma$ Eqns. \eqref{extr2} and \eqref{rege} form a closed system of non-linear algebraic equations for determination of possible extreme horizons $R_e (q, \gamma)$ and respective boundary values of dilaton $\varphi_e (q, \gamma)$.

Let us note that in the presence of an additional condition for an extremeness of horizon \eqref{extr2} the system of boundary conditions is preconditioned and asymptotic condition \eqref{bcr1} becomes ``redundant''. From the physical point of view it means that the extreme horizon does not exist for any value of mass $M_\infty $.

By exception of radical term from Eqns.\eqref{extr2} and \eqref{rege} one  can obtain an explicit dependency between dilaton mass $ \gamma$ and parameter $ \varphi_e $:
\begin {equation} \label{horg}
    R_e = \frac {1} {\gamma \sqrt {V (\varphi_e) - V \, ' (\varphi_e) /2\alpha}}.
\end {equation}
\noindent Obviously, potential of dilaton field $ V (\varphi)$ has to satisfy condition 
$$ V (\varphi_e) - \frac {1} {2\alpha} V\,'(\varphi_e) > 0. $$ 
\noindent In particular, with $\alpha = -1$, and
\begin {equation} \label {square}
    V (\varphi) = \varphi^2,
\end {equation}
the last relation becomes 
$$\varphi_e^2 +\varphi_e > 0. $$ 
\noindent In this manner horizon $R_e $ is determined for $\varphi_e \in (0, \infty)$ and
\begin {equation} \label {horg1}
    R_e = \frac {1} {\gamma \sqrt {\varphi_e^2 +\varphi_e}}\,.
\end {equation}

To compute the unknown parameter $ \varphi_e $ one should use Eq.
\begin {multline} \label {eqphi}
    \frac {1} {2\alpha} V^{\prime} (\varphi_e) \left[{1 - \frac {\gamma^2} {4} \exp{ \{-2\alpha \varphi_e \}} \frac {1} {2\alpha} V^{\prime}
     (\varphi_e)} \right] + \\
    q^2 \gamma ^2 \exp { \{ 2\alpha \varphi_e \}} \left [{V (\varphi_e) - \frac {1} {2\alpha} V^{\prime} (\varphi_e)} \right] ^2 = 0\,,
\end {multline}
following from dependencies \eqref{rege} and \eqref{horg}.

Simple notion about the qualitative behavior of solutions \eqref{eqphi} can be obtained in particular case of quadratic dilaton potential \eqref{square}. Then Eq.\ \eqref{eqphi} can be written in the form (for simplicity subscript $e$ is omitted):
\begin {equation} \label {ce}
    C (\varphi, q, \gamma) \equiv C_1 (\varphi, \gamma) - C_2 (\varphi, q, \gamma) = 0,
\end {equation}
where
$$ C_1 (\varphi, \gamma) \equiv 1 + \frac {\gamma^2} {4}\, \varphi \exp { \{2\varphi \}}, \enspace C_2(\varphi, q, \gamma) \equiv q^2 \gamma^2 \exp { \{-2 \varphi \}}\; \varphi \; (1 +\varphi)^2.$$
\begin {figure}
    \begin {center}
         \includegraphics[totalheight=5.4cm,keepaspectratio]{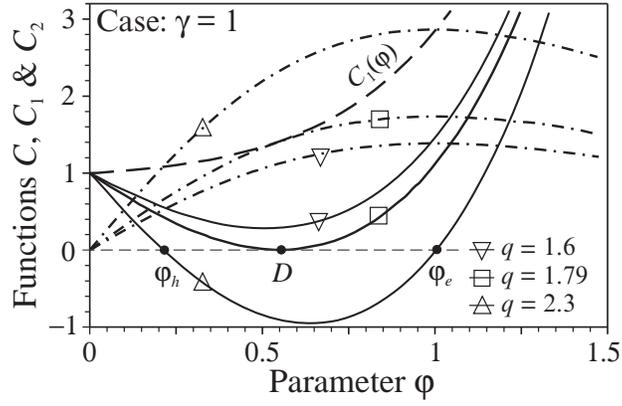} \caption {Functions $C (\varphi), C_1 (\varphi)$ and $C_2 (\varphi)$ for $ \gamma = 1$.}\label {cc1c2}
    \end {center}
\end {figure}

For given electric charge $q$ and dilaton mass $\gamma$ the function $ C_1 (\varphi)$ monotonically increases when $\varphi\in [0, \infty)$, and $C(0) = 1$ and $ \mathop {\lim}\limits_{\varphi \to \infty} C_1 (\varphi) = \infty $. By contrast, the function $ C_2 (\varphi)$ monotonically increases in interval $[0,1)$ and decreases in $(1, \infty)$. Since $C_2(0) = 0$, $C_2^{\prime}(1) = 0$ and $C_2^{\prime\prime}(1) < 0$, for $\varphi=1$ the function $C_2 (\varphi)$ has a maximum. Because of $ \mathop{\lim} \limits_{\varphi \to\infty} C_2 (\varphi) = 0$, Eq.\ \eqref{ce} can have on interval $(0, \infty)$ no more than two roots, the greater of which we shall denote $ \varphi_{e,1} $, and the smaller --- $\varphi_{e, 2}$ (hereinafter, if it does not result in misunderstanding, we shall suppose $\varphi_ {e,1} \equiv \varphi_l$ and $ \varphi_ {e, 2} \equiv \varphi_h$). These reasonings are illustrated graphically in Fig.\ \ref{cc1c2}, where the dashed line indicates the plot of function $C_1(\varphi)$, the dash-and-dot lines indicate the plots of function $ C_2 (\varphi)$ for three values $q = 1.6$, $q = 1.79$ and $q=2.3$, and the continuous lines --- the graphs of respective dependencies $C_1 (\varphi)$. In accordance with formula \eqref{horg1} the smaller root $ \varphi_h < 1$ of Eq.\ \eqref{ce} corresponds to exterior extreme horizon $R_h$ of BH and the greater root $\varphi_e$ --- to the BH with internal extreme horizon (see below the formulation of respective BVP). In order to magnitude $ \varphi_e \geq 1$ (the root is located to the right of the maximum of function $ C_2 (\varphi)$) the following condition should be satisfied 
$$ q \geq \frac {e} {2\gamma} \sqrt {1 + \frac {\gamma^2} {4} e \, ^ 2},$$
\noindent where $e\approx 2.718 \dots$ In particular, for $\gamma = 1$ we obtain $q > 2.293$ (see Fig.\ \ref{cc1c2}).

For given values $q$ and $\gamma$ it is easy to solve numerically the non-linear algebraic equation \eqref{ce} and compute the values of the roots $ \varphi_h (q, \gamma)$ and $\varphi_l (q, \gamma)$. As an example on Fig.\ \ref{fig3} the dependencies $R_h (q)$ (continuous curves) and $R_l (q)$ are shown (dotted curves) for two values of dilaton mass $ \gamma=1$ and $\gamma = 2$. Obviously when charge $q$ of exterior extreme horizon increases, $R_h$ increases linearly with coefficient $\sim 1/2$, and internal $R_l$ decreases. The points indicated through $D$ correspond to the triply degenerated horizons $R_{h} = R_l = R_d$.
\myfigures{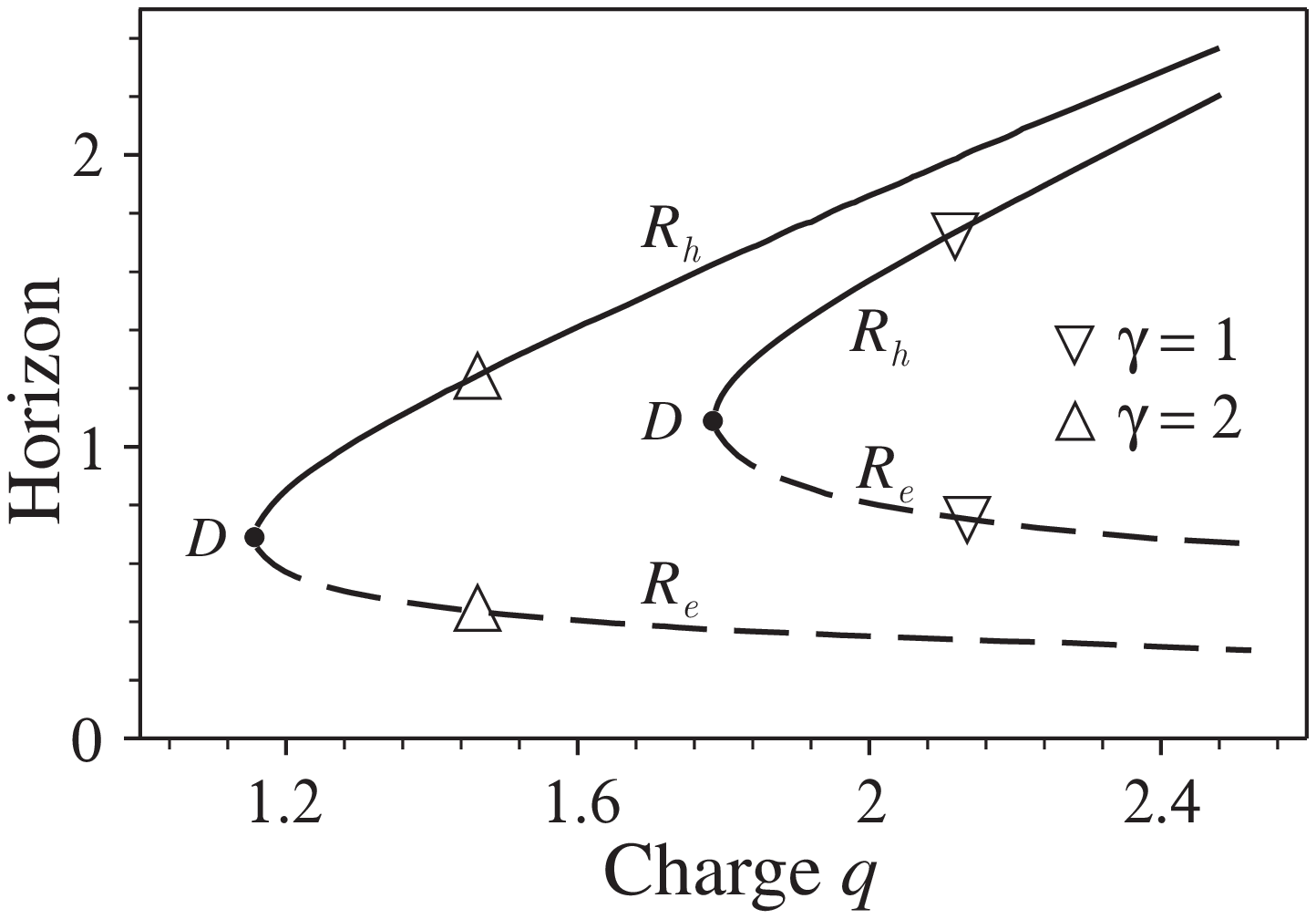}{0.46}{Relationship $R\,(q)$}{0.46}{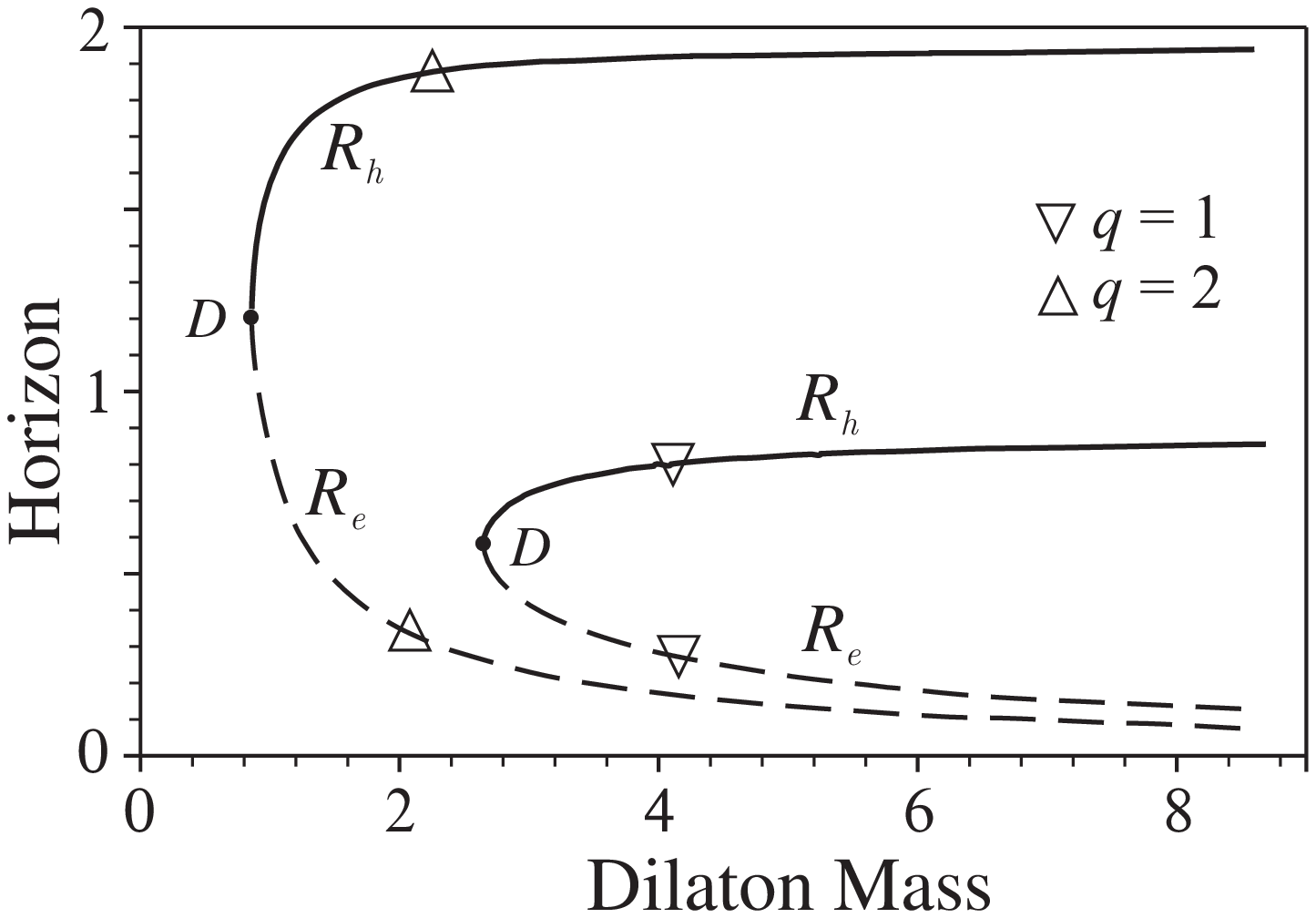}{0.45}{Relationship $R\,(\gamma)$}{0.4}

Similarly on Fig.\ \ref{fig4} the dependencies $R_h(\gamma)$ (continuous curves) and $R_l(\gamma)$ (dotted curves) for two values of a charge $q = 1$ and $q = 2$ are demonstrated. It is seen that the main variation both exterior, and internal horizons under influence of dilaton mass $ \gamma$ is localized in some neighborhood of the point of triply degeneration $D$. The large values of dilaton mass $ \gamma$ render minor influence on the magnitude of horizons.

\subsubsection* {2.2.1. BH with exterior extreme horizon} \label {extextr}

Let us suppose, that for some $q $ and $\gamma$ Eq.\ \eqref{ce} has two roots and consider the smaller of them $\varphi_h (q,\gamma)$, to which according to formula \eqref{horg1} corresponds an external horizon $R_h$. The setting of magnitude $R_h$ actually means that the left boundary of the area $D_{ext} $ is known and, therefore, system \eqref{eqns} on interval $D_{ext}$ is solved with boundary conditions \eqref{hor1} and
\begin {equation} \label {reg2}
    \varphi \, (R_h) = \varphi_{h} (q, \gamma)
\end {equation}
\noindent on the horizon, and also with dilaton asymptotic \eqref{bcr2} on the right end. The BH mass $M_{\infty} $ is obtained from the asymptotic of metric function \eqref{bcr1}.

An example of a solution obtained numerically with external extreme horizon is shown in Fig.\ \ref{fig1} (see the curve noted by $\triangledown $). In point $E $ the plot of the function $f (r)$ concerns a horizontal tangent, {\it i.e.}, the condition $f^{\prime}(R_e) =0$ is fulfilled.

For limited values of dilaton mass $\gamma$ BH has also a regular internal horizon (see below the discussion of results of the numerical experiment in Section \ref{discussion}). This horizon $R_l$ is an unknown, and, hence, in area $D_{mid} \equiv (R_l, R_h)$ it is necessary to consider a problem with free left end for Eq.\ \eqref{eqns}. Let $\{ f(r),\varphi(r), R_h \}$ be solution of the problem in the exterior area $D_{ext} $. in addition we suppose function $f(r)$ to be continuous, and function $\varphi (r)$ is smooth at point $R_h$. Then for Eq.\ \eqref{eqns} on interval $D_{mid}$ the boundary conditions look like \eqref{hor1}, \eqref{reg2} at $R_h$, and also \eqref{hor1} and \eqref{reg1} at $R_l$ (in the last two expressions one should substitute for subscript $h$ with $l$).

\subsubsection* {2.2.2. BH with internal extreme horizon} \label{intextr}

As was mentioned above in Section \ref{extrhor} the BH extreme horizon $R_l$ can be internal, {\it e.g.}, $R_{e,1} \equiv R_l < R_h$. Let us consider the formulation of BVP for Eq.\ \eqref{eqns} in this case.

Points $R_l$ and $R_h$ divide the half curve into three areas: internal $D_{int} \equiv (0, R_l)$, intermediate $D_{mid} \equiv ( R_l, R_h)$ and exterior $D_{ext} \equiv (R_h, \infty)$. Let us at first consider the problem in interval $D_{mid}$. At point $R_e$ along with the two conditions of kind \eqref{extr1} we impose an additional one

\begin {equation} \label {hori}
    \varphi \, (R_l) = \varphi_l (q, \gamma),
\end {equation}
where the value $\varphi_e (q, \gamma)$ is the greater root of Eq.\ \eqref{ce}. At the unknown right boundary $R_h$ both the condition of existence of horizon \eqref{hor1} and regularity condition \eqref{reg1} should be held. Thus the problem in the area $D_{mid} $ is self-contained.
\myfigures{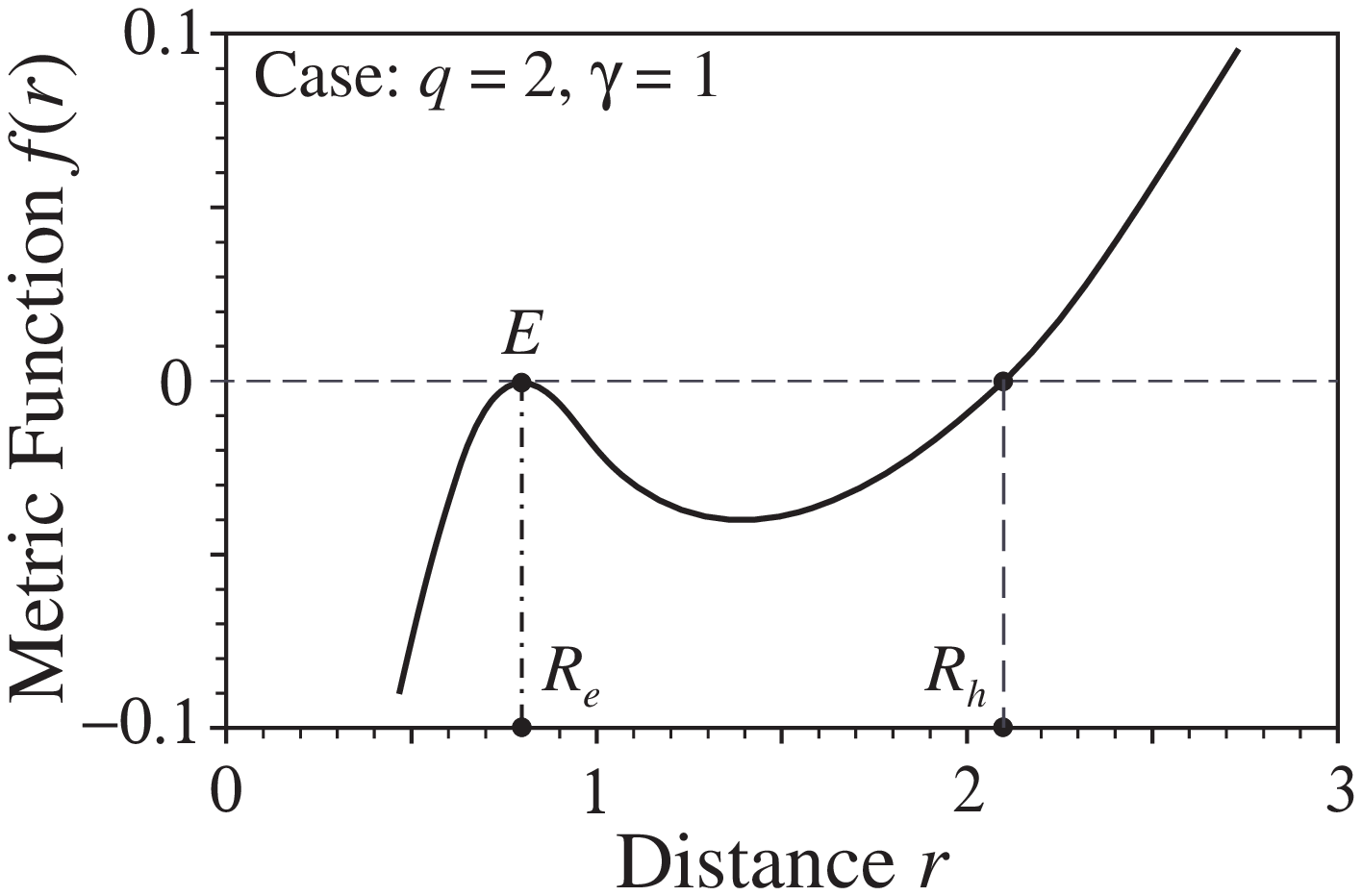}{0.45}{Solution with an internal horizon.}{0.43}{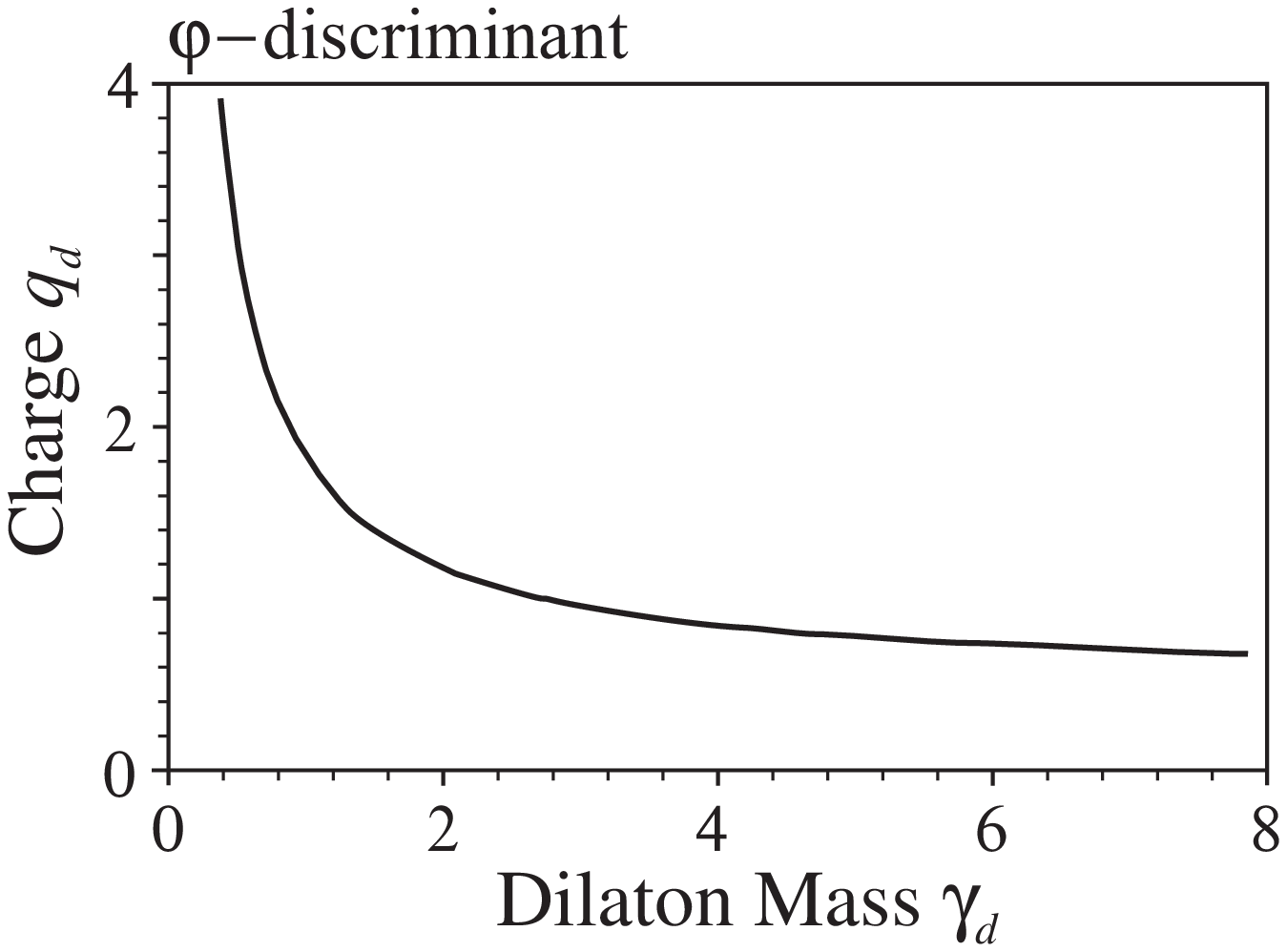}{0.43}{$\ \varphi $ --- discriminant.}{0.4}

Let the solution $\{f (r)$, $\varphi(r)$, $R_h \}$ in this area is found. Then, assuming the solutions at point $R_h$ to be continuous in area $D_{ext}$, it is necessary to solve Eqns. \eqref{eqns} with boundary conditions \eqref{hor1} and \eqref{reg2} on the left end, and also with \eqref{bcr2} on the right one.

Particular example of numerical solution $f(r)$ with internal extreme horizon $R_l \approx 0.84$ at the point $E$ for charge $q = 2$ and dilaton mass $\gamma = 1$ is presented in Fig.\ \ref{fig5}.

The formulation for seeking the solution in the internal area $D_{int}$ is similar to depicted above at the end of Subsection \ref{reg_bh}.

\subsubsection* {2.2.3. BH with triply degenerated horizon} \label {shor3}

Let for some values of charge $q$ and dilaton mass $\gamma$ Eq.\ \eqref{ce} has two roots $\varphi_h < \varphi_l$ (see Fig.\ \ref{cc1c2}). As for $\varphi \in (\varphi_h, \varphi_l)$ the inequality $C_1 (\varphi) < C_2 (\varphi)$ is satisfied, in some point $\varphi_m$ (and in view of a continuity and in some area containing point $\varphi_m$) in this interval  function $C(\varphi)$ has a negative minimum. When dilaton mass $\gamma$ decreases, the graph of function $C (\varphi, \gamma)$ rises up, the graph of function $C_1 (\varphi, \gamma)$ falls down, and the distance between roots $\varphi_h$ and $\varphi_l$ decreases. For some critical $\gamma =\gamma_d$ the minimal point $\varphi_d\equiv \varphi_m$ of function $C(\varphi,\gamma)$ concerns to horizontal axis being a single root of Eq.\ \eqref{ce}. Thus the relations
\bs  \label {phi_discr}
    \begin {align}
        C (\varphi_d, q_d, \gamma_d) &= 0, \label {degen1} \\
        \frac {\partial C} {\partial \varphi} (\varphi_d, q_d, \gamma_d) &= 0. \label {degen2}
    \end {align}
\es
take place.

Eqns. \eqref{phi_discr} determine on plain ($q, \gamma$) a smooth curve (see Fig.\ \ref{fig6}), which is called $\varphi$-discriminant of the function $C(\varphi, q, \gamma)$. The equations of discriminant in parametric form can be obtained from definition \eqref{phi_discr} \bs
    \begin {gather}
        \gamma_d = \frac {\exp(-\varphi_d)} {\varphi_d} \sqrt {2 (1 - \varphi_d)}, \label {gdeg} \\
        q_d = \frac {\exp(2\varphi_d)} {2\sqrt {1-\varphi_d^2}}. \label {qdeg}
    \end {gather}
\es 
The simple calculations demonstrate that relation \eqref{degen2} expresses the condition of a vanishing of second derivative of metric function $f (r)$ in horizon $R_d$
$$ f^{\prime\prime} (R_d) = 0\,.$$ 
Such horizon is called in articles \cite {jinr01_221, jinr02_1} triply degenerated.

Thus, the algorithm for solving the problem for BH with triply degenerated horizon looks like the following. It is more convenient to set dilaton mass $\gamma =\gamma_d$. Then according to formula \eqref{gdeg} it is possible to compute $0 < \varphi_d (\gamma) < 1$, and by means of relationship \eqref{horg1} -- the horizon $R_d$ of BH. The BH charge $q$ is found from \eqref{qdeg}. Therefore, in external $D_{ext} $ /internal $D_{int} $ domain BVP for triply degenerated horizon seems to be a problem with fixed left/right boundary $R_d$, where one sets conditions of kind \eqref{hor1} and \eqref{reg2}. If the solution of this problem is found, then respective BH mass $M_{\infty} $ can be found from the asymptotic expression \eqref{bcr1}.
\begin{figure}
    \begin{center}
        \includegraphics[totalheight=5cm,keepaspectratio]{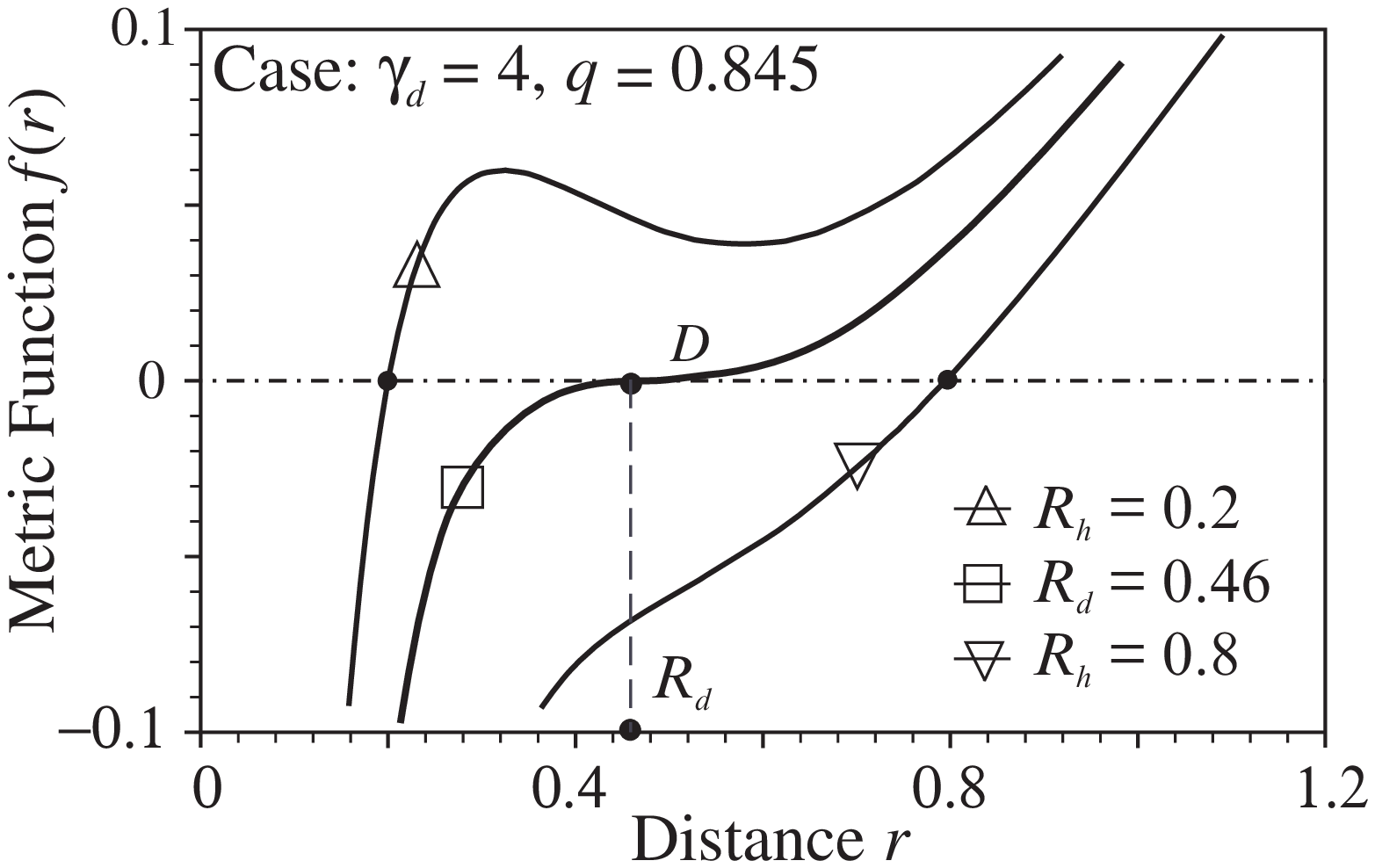} \caption{An example of a solution $f (r)$ with triply degenerated horizon.}\label{fig7}
    \end{center}
\end{figure}

An example of triply degenerated solution $f(r)$ obtained numerically for $q_d \approx 0.836$ and $\gamma_d = 4$ is presented in Fig.\ \ref{fig7}. Point $D$ ($R_d \approx 0.456$) is an inflection point for metric function $f (r)$, {\it i.e.}, in addition with $f (R_d) = 0$ the conditions $f^{\prime} (R_d) = 0$ and $f^{\prime\prime} (R_d) = 0$ are fulfilled also.

\section{Numerical Method for Solving a Problem  with Fixed Boundary} \label {fixed}

As was mentioned above, in a case of BH with exterior extreme horizon $R_h$ it is necessary to decide a BVP for Eq.\ \eqref{eqns} in domain $D_ {ext} $ with fixed physical parameters at the right end $R_h$ and system with three boundary conditions \eqref{hor1}, \eqref{reg2} and \eqref{bcr2}. For the numerical implementation it is expedient to apply the continuous analogue of Newton method (CANM) which  is very suitable for treating of numerous problems in physics \cite{pazppsl_99}.

Let us introduce an ``actual'' infinity $R_\infty < \infty $ and consider the non-linear functional equation
\begin {equation} \label {eqy}
    \chi (y) = 0
\end {equation}
\noindent with respect to pair $y \equiv \{f, \varphi \} $, $y \in
C^1 [R_h, R_\infty] \times C^2 [R_h, R_\infty] $, where the
coordinates $\chi^n (y)$, $n=1,2, \dots, 5 $, are determined as
follows: \bs
    \begin {eqnarray}
        \chi ^ {(1)} &\equiv& -f \,' + F (r, f, \varphi, \varphi\,'), \\
        \chi ^ {(2)} &\equiv& -f \left (\varphi\,'' - \frac {1} {r} \varphi\,' \right) + \Phi (r, \varphi, \varphi\,'), \\
        \chi ^ {(3)} &\equiv& f (R_h), \\
        \chi ^ {(4)} &\equiv& \varphi (R_h) - \varphi_h (q, \gamma), \label {coord4} \\
        \chi ^ {(5)} &\equiv& \varphi (R_\infty) + \frac {\alpha q^2} {\gamma^2 R^4_{\infty}}.
    \end {eqnarray}
\es

Let Eq.\ \eqref{eqy} have an insulated solution $y ^ {*} \equiv \{f^*(r), \varphi^*(r) \}$ and initial approximation $y^{0} \equiv \{f^0 (r), \varphi^0 (r) \} $ near to solution $y^*$ is known. For small values of charge $q $ and dilaton mass $\gamma$ we can use as initial approximation asymptotics \eqref{bcr1} and \eqref{bcr2}.

Let us introduce continuous parameter $t\in [0, \infty)$ and assume trajectory $y (t)$ satisfying the abstract Cauchy problem
\begin {equation} \label {bhevol}
    \chi ' (y (t)) \frac {d \, y (t)} {d \, t} + \chi (y (t)) = 0, \quad y (0) = y_0.
\end {equation}

Here $\chi^{\prime}(y)$ is Frechet's derivative of the non-linear operator $\chi(y)$. In \cite {zhmp_73} it is shown that when the operator $\chi (y)$ is smooth in some vicinity of the sought solution $y^*$ the limit relationship
$$\lim \| y (t) -y^*\| \xrightarrow [{t \to \infty}] {} 0\,.$$
takes place.

For the numerical solution of Cauchy problem \eqref{bhevol} it is easy to take advantage of the explicit Euler method on non-uniform grid $t_ {k+1} = t_k + \tau_k$ with variable step $\tau_k$, $k=0,1, \dots$ As a result we come to iterative process
\begin {eqnarray}
    \chi\,'(y_k) \: w_k &=& -\chi (y_k)\,, \label {bhiter} \\
    y_{k+1} &=& y_k + \tau_k\: w_k\,, \label {tare}
\end {eqnarray}
\noindent which allows on each iteration $k $ by help \eqref
{bhiter} to compute next approximation $y_{k+1}$ to the exact
solution. When $\tau_k=1$ we come to the classic Newton method.

In \cite {zhmp_73} the convergence of iterations \eqref{bhiter}, \eqref{tare} to the solution of the evolution Cauchy problem \eqref{bhevol} in finite time interval $t$ when $\tau_k \to 0$ is proved, if in vicinity of exact solution $y^{*}$ operator $\chi (y)$ satisfies some additional conditions.

The variation of step $\tau_k$ can be ruled during the iteration process \cite {pazppsl_99, ek_81}.

On each iteration Eq.\ \eqref{bhiter} in the problem under
consideration is equivalent to the following linear problem for
the coordinates of correction vector $w (r) \equiv \{\xi (r),
\eta(r) \} $ (for simplicity hereinafter index $k $ will be
omitted): \bs \label {canm1}
        \begin {gather}
            -\xi \, ' + \frac {{\partial F}} {{\partial \varphi \, '}}\eta \, ' +
            \frac {{\partial F}} {{\partial f}} \xi + \frac {{\partial F}} {{\partial \varphi}} \eta             f ' - F\left (r, f, \varphi, \varphi \, ' \right), \\
            \begin {split}
        -f\left ({\eta \,'' + \frac {1} {r} \eta \, '}\right) + \frac {\partial \Phi} {\partial\varphi \, '}\eta \, ' - \left ({\varphi \,'' + \frac {1} {r} \varphi \, '}\right) \xi &+ \frac {\partial \Phi} {\partial \varphi} \eta = \\
         &f\left ({\varphi \,'' + \frac {1} {r} \varphi \, '} \right) - \Phi \left ({r, \varphi, \varphi \, '} \right) \,
            \end {split} \\
        \xi (R_h) = -f (R_h), \label {bcxi} \\
        \eta (R_h) = \varphi_h (q, \gamma) -\varphi (R_h), \label {bceta1} \\
         \eta(R_\infty)=-\varphi(R_\infty)-\frac{\alpha q^2} {\gamma^2 R_\infty^4} . \label {bceta2}
    \end {gather}
\es It is convenient to select initial approximation $\{f_0 (r),
\varphi_0 (r) \} $ satisfying initial conditions \eqref{bcr2},
\eqref{hor1} and \eqref{reg2}. Then on each iteration right hand
sides \eqref{bcxi}--\eqref{bceta2} are equal to zero, and the
contribution of boundary conditions in error $\delta (\tau_k)$
will be zero, respectively.

Let us suppose that $\{\xi (r), \eta (r) \} $ is a solution of the problem \eqref{canm1}. Then the next approximation to the exact solution is computed from relation (see formula \eqref{tare})
\begin {equation}
    f^{k+1} = f^k + \tau_k\, \xi^k, \quad \varphi^{k+1} = \varphi^k + \tau_k \, \eta^k.
\end {equation}

In cases, when horizon $R_h < R_{e, 1}$ or $R_h > R_{e, 2}$, dependence $R_h (M_\infty)$ appears to be univalent (see below the discussion in Section \ref{discussion}). This allows to consider the quantity $R_h$ as a parameter and thus substitute the problem with free left boundary problem with fixed boundary $R_h$. In this case the right hand side of the expression \eqref{coord4} becomes form similar to Eq.\ \eqref{reg1}:
\begin {equation*}
    \chi ^ {(4)} \equiv \Phi (R_h, \varphi_h, \varphi \, '_h) \,
\end {equation*}
and boundary conditions \eqref{bceta1} is substituted by
\begin {equation*} 
    \left. {\frac {{\partial \Phi}} {{\partial \varphi}}} \right |_{R_h} \eta (R_h) +
    \left. {\frac {{\partial \Phi}} {{\partial \varphi \, '}}} \right |_{R_h} \eta \, ' (R_h)      - \Phi \left ({R_h, \varphi_h, \varphi \, '_h} \right).
\end {equation*}

On each iteration $k$ a linear BVP \eqref{canm1} is solved in finite interval $\left(R_h, R_\infty\right)$. For discretization the method of collocation at the Gaussian knots of grid exponentially condensed to horizon $R_h$ is used.

Let $u^*$ be the exact solution of the original continuous problem \eqref{eqy} on a finite interval of time, $u_h^{*}$ --- the exact solution corresponding to the  discretized non-linear problem
\begin {equation} \label {grideq}
    \chi_h (u_h) = 0
\end {equation}
in a finite interval of time, $u_h^k $ -- approximation to $u_h$ after $k$th iteration when condition $\| \chi_h (u_h^k) \| \leq \varepsilon$, where $0 < \varepsilon \ll 1$, is fulfilled. Let us note that if the discretization method does not vary from iteration to iteration, then the grid representations for \eqref{canm1} follow from grid representation \eqref{grideq} of original equation \eqref{eqy}.

For convergence estimate of the method we shall consider the inequality
$$\| y^* - u_h^k \|_h \leq \|y^* - u^* \|_h + \|u^* - u_h \|_h + \|u_h - u_h^k \|_h\,.$$ 
It is possible to show \cite{pazppsl_99} that $\|u^*-u_h \|_h \leq O(h^r)$, $\|u_h ^* - u_h^k \| \leq B_h \| \chi_h (u_h^k) \|$, where $B_h$ -- some constant. Then for $B_h \ | \chi_h (u_h^k) \| \ll O(h^r)$, which is fulfilled for enough small $\varepsilon$, the precision of obtained approximate solution is specified by two first summands on the right hand side in the inequality.

The error $\delta_\infty = \|y ^*-u ^* \|_ h $ is investigated numerically on a fixed grid for different values of actual infinity $R_\infty $, which one select, so that $\delta_\infty$ remained small with respect to two other summands. Thus, the precision of the approximated solution is close to the theoretical estimate of method of difference approximation of Eq.\ \eqref{eqy}.

\section {Numerical Method for Free-Boundary Problem}\label {freebound}

An exhausting enough review of methods for solving free-boundary problems is presented in treatise \cite{vab_87}. In present work the boundary-value problem for Eqns.\ \eqref{eqns} with free external $R_h$ and given extremal internal $R_l$ horizons is rendered to non-linear eigenvalue problem with spectral parameter $R_h$, which in turn is solved by CANM. Let us remark that such approach has good reputation for treating various problems in astrophysics \cite{btfy01, btfy02}, the theory of Josephson junctions \cite{bt02}, {\it etc.}

A formal shortcoming for solving the problem with unknown external horizon by CANM is the absence of explicit dependence between the equations and boundary conditions and horizon $R_h$. In order to enter explicitly parameter $R_h$ we introduce new variable $x$ under the formula
\begin {equation} \label {transf}
    x = \frac {r-R_l}{R_h-R_l}\,.
\end {equation}

After the change of variables \eqref{transf} interval $ [R_l, R_h]$ renders to $[0, 1]$, at that $d/dr = (R_h-R_l) ^ {-1} d/dx$. Let us assume $z\equiv \{y, R_h \} $, $y\equiv \{f, \varphi \}$, $z \in C^1 [R_h, R_\infty] \times C^2 [R_h, R_\infty] \times \mathbb {R}$. Then BVP for Eq.\ \eqref{eqns} can be written similarly to Eq.\ \eqref{eqy}: \bs
    \begin {eqnarray}
         \chi(y,R_h)&=&0, \label {eqz} \\
        N (y) &=& 0\,, \label {normz}
    \end {eqnarray}
\es 
where vector $\chi (z)$ is specified using expressions (to avoid the introducing of redundand notations further in this section we substitute $ (.)^{\prime} \equiv d (.) /dx $) 
\bs \label {freeb}
    \begin {eqnarray}
        \chi ^ {(1)} &\equiv& -f \, ' + \bar {F} \left [x, f, \varphi, \varphi \, ', R_h \right] =0, \label {metr} \\
        \chi ^ {(2)} &\equiv& -f\left (\varphi \,'' + \frac {1} {x} \varphi \, ' \right) + \bar {\Phi} \left [x, \varphi, \varphi \, ', R_h) \right] = 0, \label {dil} \\
        \chi ^ {(3)} &\equiv& f (0), \\
        \chi ^ {(4)} &\equiv& f (1), \\
        \chi ^ {(5)} &\equiv& \bar {\Phi} (1, \varphi_h, \varphi \, '_h, R_h),
    \end {eqnarray}
\es and the left part of ``norm condition'' \eqref{normz} has the form
\begin {equation} \label {nrmr}
    N (y) \equiv \varphi (0) - \varphi_l (q, \gamma).
\end {equation}
Here through $\bar {F}$ and $\bar {\Phi}$ the right hands of the BH Eqs.\ \eqref{eqns} after substitution in them \eqref{transf} are indicated:
\begin {eqnarray*}
    \bar {F} &\equiv& (R_h-R_l) \, F \left [r(x, R_h), f, \varphi, (R_h-R_l)^{-1} \varphi\,' \right], \\
    \bar {\Phi} &\equiv& (R_h-R_l) ^2 \, \Phi \left [r (x, R_h), \varphi, (R_h-R_l)^{-1} \varphi\,' \right].
\end {eqnarray*}

As in Section \ref{fixed}, we introduce continuous parameter $t \in [0, \infty)$, supposing the CANM relation being
\bs \label {cauchi}
    \begin {gather}
        \chi\,'_y (y, R_h)\, w + \chi\,'_ {R_h} (y, R_h) \, \rho + \chi (y, R_h) = 0, \label {chi} \\
        N\,' (y) \, w + N (y) =0, \label {nrm} \\
        w = \dot {y}, \quad \rho = \dot {R_h}. \label {bhdiff}
    \end {gather}
\es 
Problem \eqref{chi} -- \eqref{bhdiff} has to solved with initial conditions
\begin {equation} \label {icond}
    y (0) = y_0, \quad R_h (0) = R_{h, 0}.
\end {equation}

The numerical realization of CANM can be implemented by different methods for approximated solving of differential equations. Let us consider further Euler iteration process corresponding to Cauchy problem \eqref{cauchi} on non-regular generally grid $t_{k+1} = t_k + \tau_k $, $k = 0, 1, \ldots $. At $k$-th  iterations it is necessary to solve the linear operator equations \eqref{chi} and \eqref{nrm}, after that the subsequent approximation to the exact solution is found through the formulas, following from relations \eqref{bhdiff}:
\begin {equation} \label {bhiter2}
    y^{k+1} = y^k + \tau_k\, w^k, \quad R_h^{k+1} = R_h^k + \tau_k\, \rho^k.
\end {equation}

We shall search the solution of linear equation \eqref{chi} in the form (for simplicity henceforth we omit iteration index $k$)
$$ w = u + \rho\, v\,,$$
where $u(x)$ and $v(x)$ are new unknown functions. Substituting this decomposition into \eqref{chi} and equating coefficients we receive the system 
\bs \label {lineq}
    \begin {eqnarray}
        \chi\,'_ y (y, R_h) \, u = -\chi (y, R_h), \label {ueq} \\
        \chi\,'_ y (y, R_h) \, v = -\chi\,'_ {R_h} (y, R_h). \label {veq}
    \end {eqnarray}
\es 

Let the solutions of these equations be computed. Then the derivative $\rho $ can be obtained from \eqref{nrm}
\begin {equation} \label {ro}
    \rho = - \left [N \, ' (y) v\right] ^ {-1} \left [N (y) + N \, ' (y) u\right] .
\end {equation}
It is expedient to select the initial approximation $y_0$ so that it obeys norm condition \eqref{nrm}. In that case expression
\eqref{ro} is simplified:
\begin {equation} \label {ro1}
    \rho = - \left [N \, ' (y) v\right] ^ {-1} N \, ' (y) \, u .
\end {equation}
\myfigures{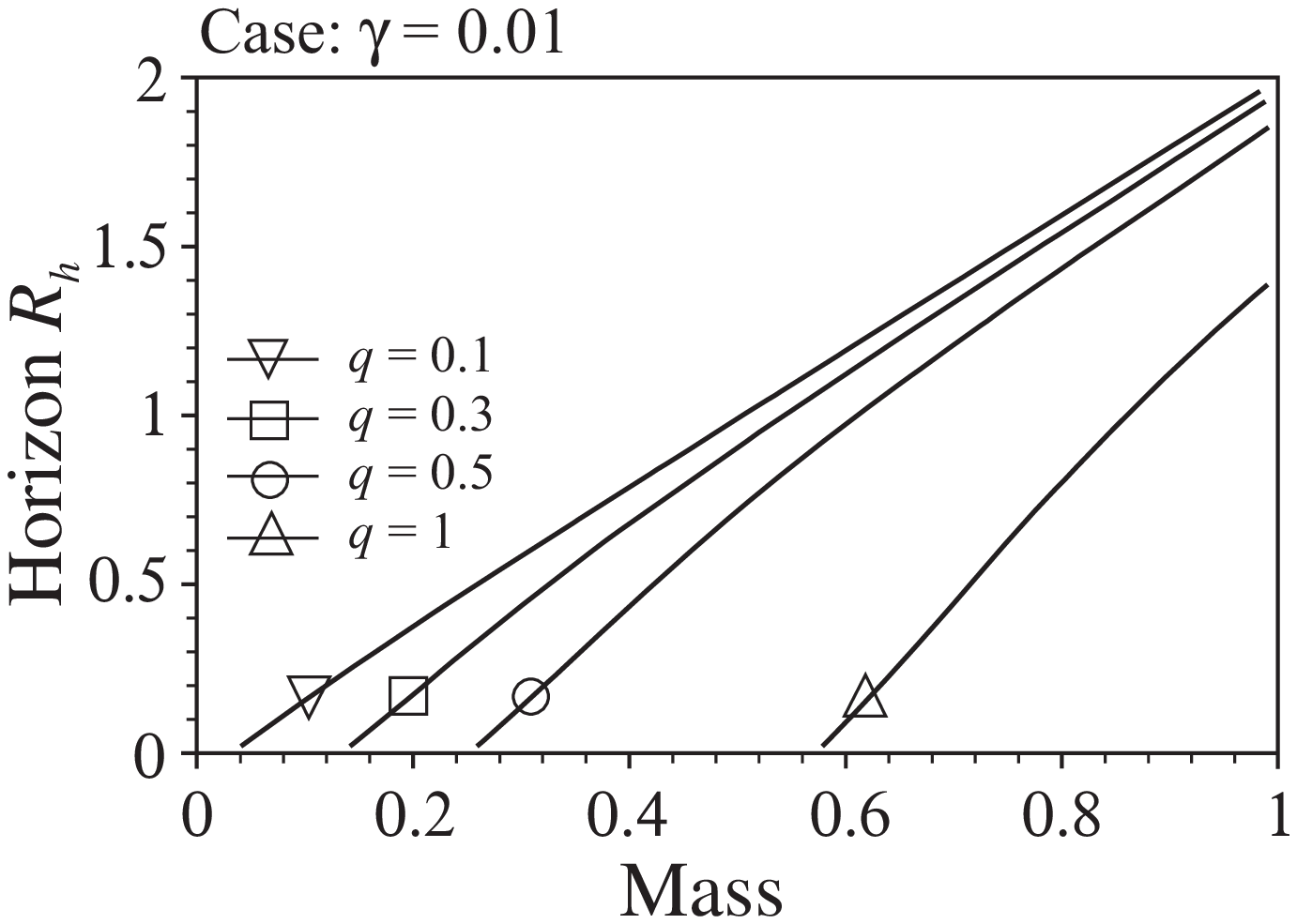}{0.46}{Dependence $R_h (M_\infty)$ for small $\gamma$.}{0.48}{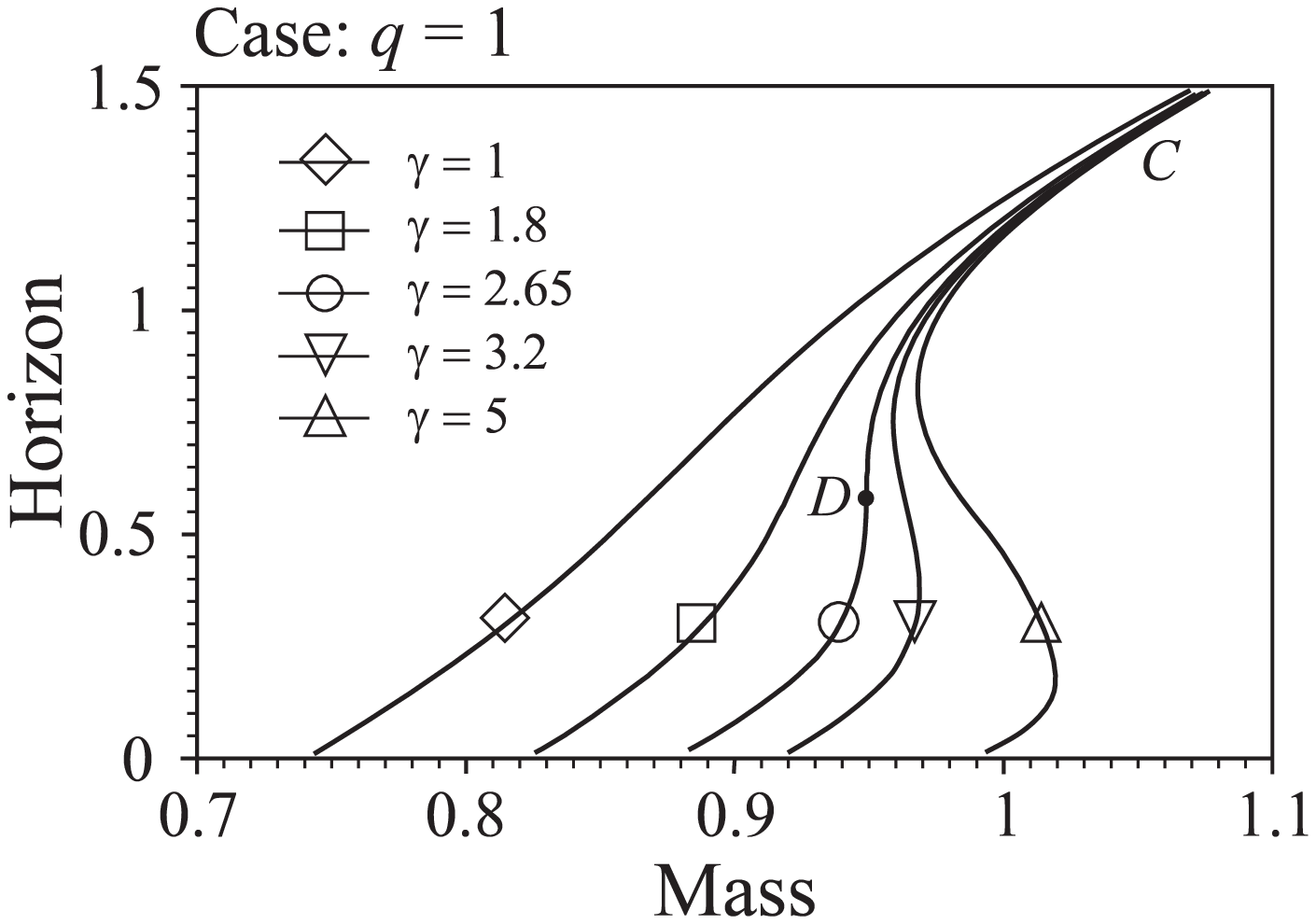} {0.45} {Dependence $R_h (M_\infty)$.}{0.4}

Let us give the explicit form of Eqs.\eqref{lineq} and \eqref{ro}. Let $u \equiv \{u_f (x), u_\varphi (x) \} $, and $v \equiv \{v_f (x), v_\varphi (x) \}$. Then we have
\bs \label {canm2}
    \begin {gather}
        -u \, '_f + \frac {\partial \bar {F}} {\partial \varphi \, '} u_\varphi \, ' + \frac {\partial \bar {F}} {\partial f} u_f + \frac {\partial \bar {F}} {\partial \varphi} u_\varphi = f \, ' - \bar {F} \left (x, f, \varphi, \varphi \, ', R_h \right), \\
        \begin {split}
        -f\left ({u_\varphi \,'' + \frac {1} {r} u_\varphi \, '} \right) +
        \frac {\partial \bar {\Phi}} {\partial \varphi \, '} u_\varphi \, ' - \left ({\varphi \,'' + \frac {1} {r} \varphi \, '} \right) u_f + \frac {{\partial \bar {\Phi}}} {{\partial \varphi}} u_\varphi &= \\
        f\left ({\varphi \,'' + \frac {1} {r} \varphi \, '} \right) - \bar {\Phi} \left (x, \varphi, \varphi \, ', R_h \right),
        \end {split} \\
        u_f (0) = -f (0), \\
        u_f (1) = -f (1), \\
        \left. {\frac {\partial \bar \Phi} {\partial \varphi}} \right |_{x = 1} u_\varphi (1) + \left. {\frac {\partial \bar \Phi} {\partial \varphi \, '}}\right |_{x = 1} u '_\varphi (1) = -\bar \Phi \left ({1, \varphi, \varphi \, ', R_h} \right),
    \end {gather}
\es \bs \label {canm4}
    \begin {gather}
        -v \, '_f + \frac {\partial \bar {F}} {\partial \varphi '} v_\varphi \, ' +
        \frac {\partial \bar {F}} {\partial f} v_f + \frac {\partial \bar {F}} {\partial \varphi} v_\varphi         -\frac {\partial \bar {F}} {\partial R_h} \, \\
        -f\left ({v_\varphi \,'' + \frac {1} {r} v_\varphi \, '} \right) + \frac {\partial \bar {\Phi}} {\partial \varphi \, '} \, v_\varphi \, ' - \left ({\varphi \,'' + \frac {1} {r} \varphi \, '} \right) v_f + \frac {\partial \bar {\Phi}} {\partial \varphi} v_\varphi = -\frac {\partial \bar {\Phi}} {\partial R_h} \, \\
        v_f (0) = 0 \,, \\
        v_f (1) = 0 \,, \\
        \left. {\frac {\partial \bar\Phi} {\partial \varphi}} \right |_{x = 1} v_\varphi (1) +
        \left. {\frac {\partial \bar\Phi} {\partial \varphi \, '}} \right |_{x = 1} v '_ \varphi (1)         \left. -\frac {\partial \bar \Phi} {\partial R_h} \right |_{x = 1} .
    \end {gather}
\es
\begin {equation} \label {rophi}
    \rho = \frac {\varphi (q, \gamma) -\varphi (0) -u_\varphi (0)} {v_\varphi (0)}\,.
\end {equation}

Thus, the iteration process is conducted in following order. For given initial approximation $f_0 (x)$, $\varphi_0(x)$ and $R_h$ we compute increments $u(x)$, $v(x)$ solving boundary problems \eqref{canm2} and \eqref{canm4}. Further by formula \eqref{rophi} we find increment $\rho $ of horizon $R_h$. The next approximation of the exact solution is obtained using \eqref{bhiter2}.

The linear BVPs \eqref{canm2} and \eqref{canm4} are solved numerically through the collocation method of order $O(h^4)$ in the Gaussian knots of grid, which is exponentially condensed to horizon $R_h$ \cite{btfy01}. Let us note that the left-hand sides both the equations and the boundary conditions are identical, what simplifies the solution of respective matrix problem.

Let the solution in domain $D_{mid}$ be found. Then supposing a continuity of functions at point $R_h$, the problem in external domain $D_{ext}$ is treated as a problem with fixed left boundary (Section \ref{fixed}). If it is necessary to find the solution in internal domain $D_{int}$ (see the end of Section \ref{blhol1}) the corresponding BVP with the free left boundary can be solved as is depicted above.

Apparently the solution of the problem with extremal external and regular internal horizons can be obtained by methods presented in Sections \ref{fixed} and \ref{freebound}.

\section {Discussion of Numerical Results} \label {discussion}

>From physical point of view very relevant is the link between BH mass $M_\infty $ and horizon $R_h$ for different values of parameters $q$ and $\gamma$.

The numerically obtained dependence $R_h (M_\infty)$ for  small dilaton mass $\gamma=0.01$ and $q \leq 1$ is demonstrated in Fig.\ \ref{fig8}. For such values of $\gamma$ it is necessary to expect that uniqueness of the dependence $R\,(M_\infty)$ and linearity for large values $M_\infty$ are saved in a wide range of variation of charge $q$. This deduction is similar to the recent results \cite{tt00, tt01} concerning the case of massless dilaton. It is easy to calculate that the triple degeneration of horizon, which initiates an essential influence of mass $\gamma$ on number and form of horizons (see below), takes place for $q > 130$. In this sense the curves in Fig.\ \ref{fig8} have test nature.

The influence of finite dilaton mass $\gamma$ is shown in Fig.\ \ref{fig9} for given charge $q=1$. When $\gamma < \gamma_d \approx 2.65$ (see curves indicated through $\lozenge$ and $\Box$) dependencies $R (M_\infty)$, though hardly distorted, remain single-valued. Curve $R (M_\infty)$ (denoted through $\bigcirc$) corresponding to extremal value $\gamma_d$, has in the point of triple degeneration of horizon $D$ ($R_d \approx 0.58$) vertical tangent (for inverse function $M_\infty (R_h)$ point $D$ is an inflection point). The corresponding solution for metric function $f (r)$ is plotted in Fig.\ \ref{fig7}, curve $\square$.

For further increase of dilaton mass $\gamma > \gamma_d$ BH has (see Section \ref{extrhor}) two extremal horizons $R_l (q, \gamma)$ and $R_h (q, \gamma)$, $R_h > R_l$. In the graph of dependence $R \, (M_\infty)$ the presence of horizons $R_l$ and $R_h$ expresses in appearance of typical $S$-shaped curves (in given case indicated through $\triangledown $ and $\triangle $). Two different solutions of Eq.\ \eqref{eqns} with different masses $M_h (q, \gamma)$ and $M_l (q, \gamma)$ correspond to these horizons.
\begin{figure}
    \begin{center}
        \includegraphics[totalheight=5.4cm,keepaspectratio]{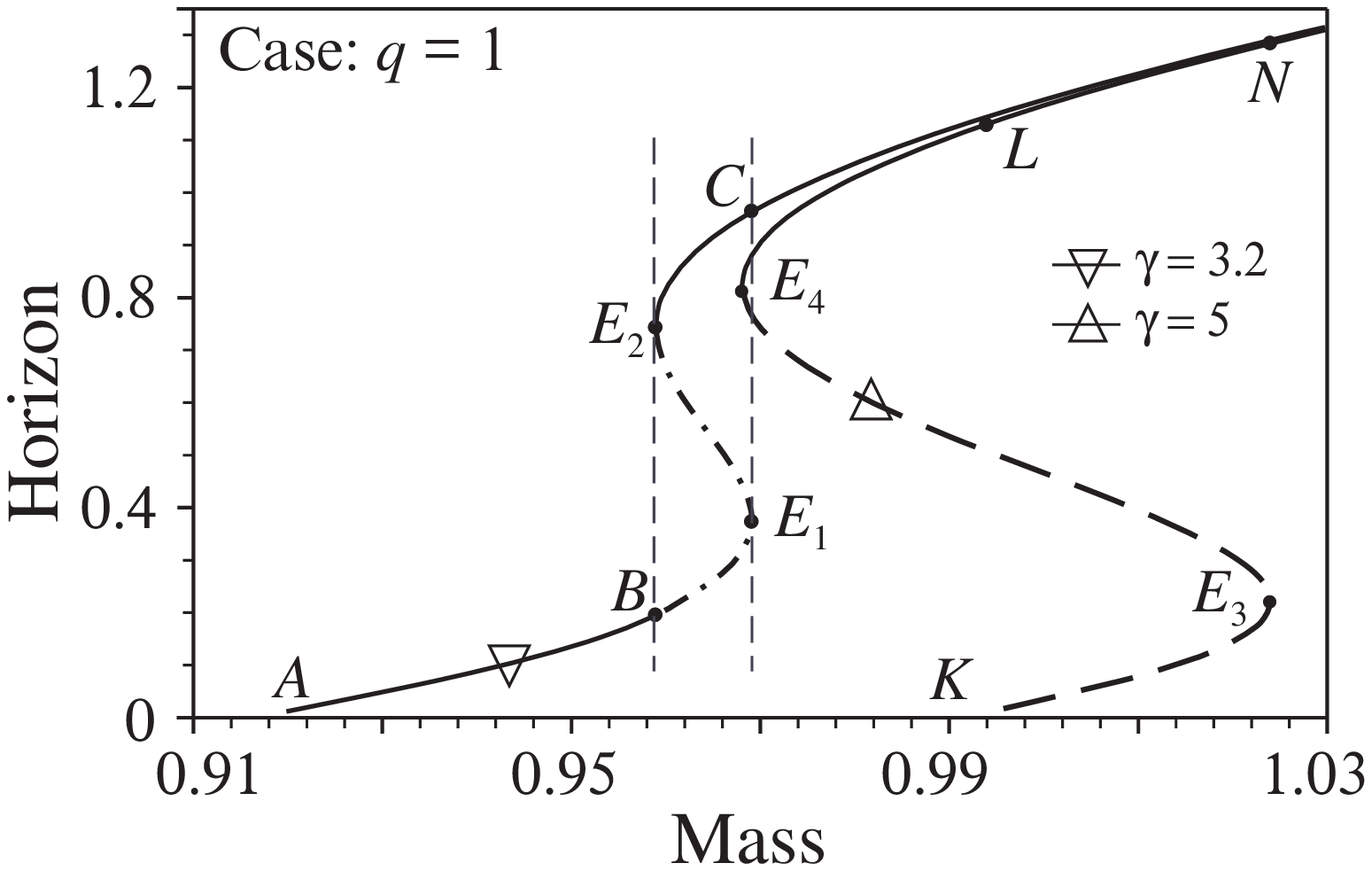} \caption{``Formation'' of horizons in BH.} \label{scurve}
    \end{center}
\end{figure}

Let us fix the value $\gamma = 3.2 > \gamma_d$ (curve noted by symbol $\triangledown $ in Fig.\ \ref{fig9}). In Fig.\ \ref{scurve} this curve together with curve $\triangle $ are plotted in an appropriate scale. Moving on arc $ABE_1E_2C $ from left to right the dynamics of variation of number of horizons apparently looks like the following. From point $A $ up to point $B_1$ BH has single regular horizon $R_h$, on vertical straight $BE_2$ -- two horizons, and horizon in point $B $ is regular, and in point $E_2$ -- external extremal one ( $R_h \approx 0.72$). Further, for given $M_\infty$ on the segment between vertical straights $BE_2$ and $E_1C $ BH has three regular horizons, on the straight line $E_1C $ -- two, and vertical coordinate $R_l \approx 0.42$ of point $E_1$ is an internal extremal horizon. At last, on the right of point $C $ BH has again only one regular horizon ($R = R_h$).

The ``motion'' of extremal horizons $R_l$ and $R_h$ as a result of the variation of mass $\gamma$ for values of electric charge $q = 1$ and $q = 2$ is presented in Fig.\ \ref{fig4}. Similar dependence of the magnitude of horizons from  charge $q $ for fixed $\gamma = 1$ and $\gamma = 2$ is demonstrated in Fig.\ \ref{fig3}.

For large enough $\gamma$ the vertical coordinate $R_h$ of point $E_2$ increases linearly (see Fig.\ \ref{fig4}), and the coordinate $R_l$ of point $E_1$ comes downward. Thus the point $M_0$ (mass specified by the electrical field and the dilaton), fulfilling the formal equality $R \, (M_0) = 0$, can be located on the right to the vertical straight line $BE_2$, {\it i.e.}, will be executed $M_0 > M_h$ (see the curve noted by $\triangle $ in Fig.\ \ref{scurve}). This means, that for respective $\gamma$ (in particular case $\gamma = 5 $) the distribution of horizons is as follows: at point $E_2\;$ BH has a single extremal horizon (see Section \ref{extextr}). Further, in the domain between vertical straights through $E_2$ and $M_0$ BH has two regular horizons. In the right of the point $M_0$ up to $E_1$ BH has three regular horizons, at point $E_1$ -- two horizons $R_l$ and $R_c = R_h$, and internal $R_l$ is extremal. At last, for large enough masses $M_\infty$ (in the right of point $E_1$) BH has a single regular horizon.
\myfigures{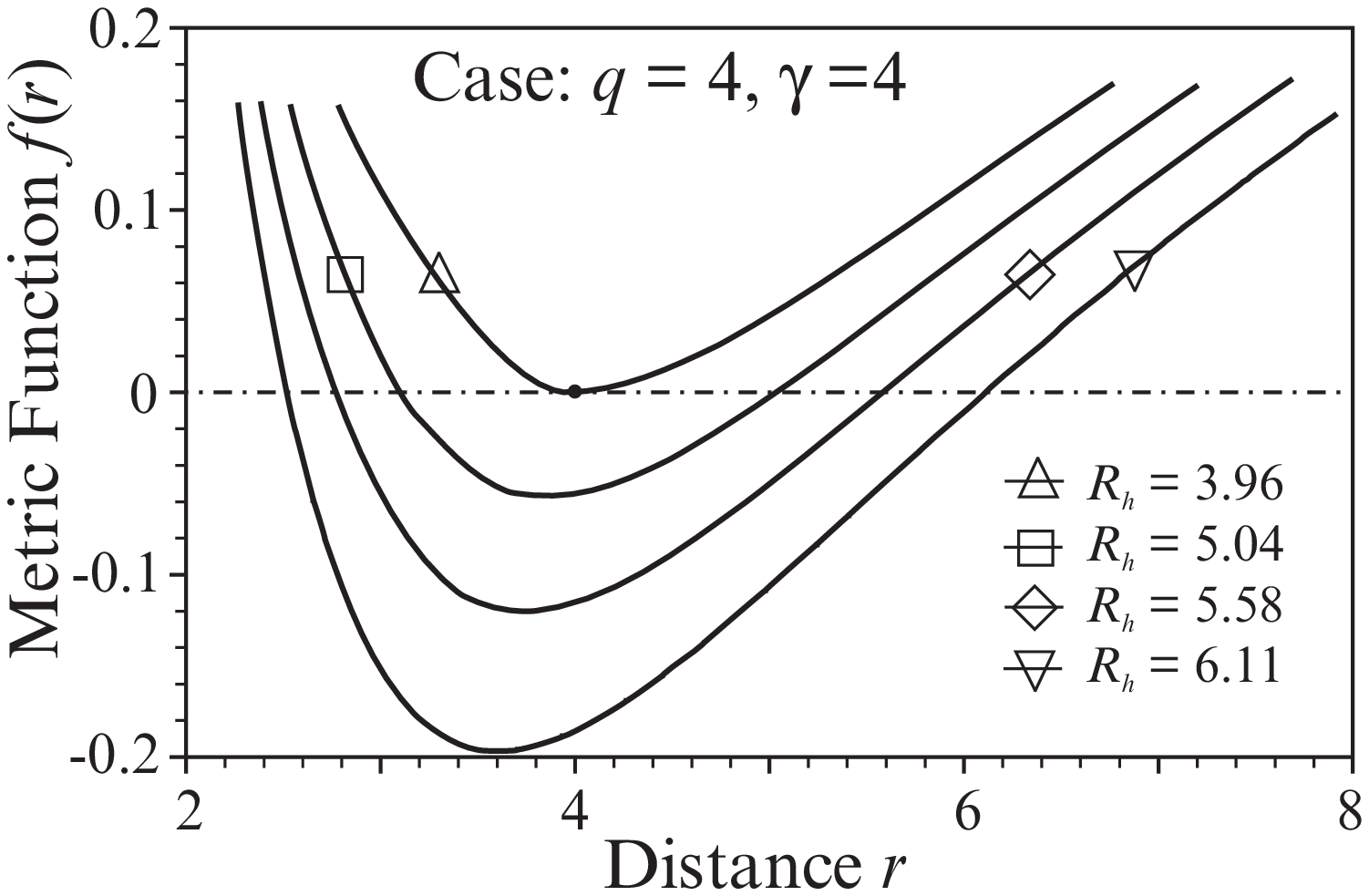}{0.46}{An example of solution $f (r)$ with two regular horizons.}{0.46}{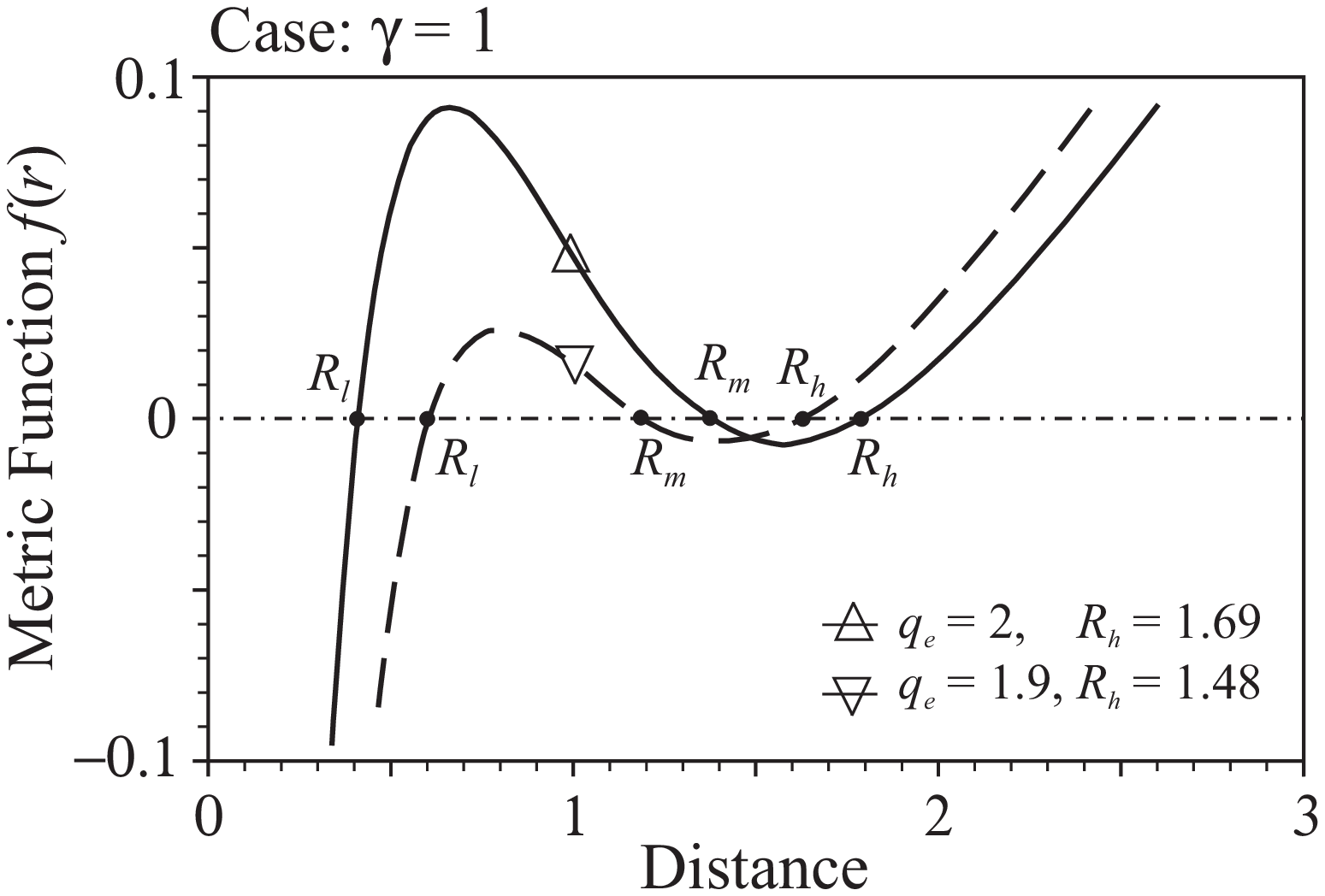}{0.45}{Examples of solutions $f (r)$ with three regular horizons.}{0.4}

In Fig.\ \ref{fig11} the examples of solutions with two regular horizons in neighborhood of extremal solution $\bigtriangleup$, $M_\infty \approx 3.96$ (curves noted by characters $\square$, $\lozenge$ and $\triangledown$) are plotted. At that computation values of the BH mass: $\square - M\infty = 4.1$, $\lozenge - M\infty = 4.22$ and $\triangledown - M_\infty = 4.36$.

In Fig.\ \ref{fig12} two examples of solutions with three horizons for $\gamma=1$, $q=1.9 $, $R_h \approx 1.48 $ (the dotted curve noted by a character $\triangledown $) and $q=2$, $R_h \approx 1.69 $ (continuous curve $\triangle $) are shown.

It is easy to obtain a condition, for which on the leftward of vertical straight line $BE_2$ through point $R_h$ (see Fig.\ \ref{scurve}) BH does not have any horizons. For this purpose we mark by $M_l$ and $M_h$ the masses of two BH, for which $R_l$ and $R_h$ are respectively internal and external extremal horizons. Let us build Hermitian cubic polynomial $S_3 (R)$ interpolated the function $M_\infty (R)$ on the interval $R\in [R_l, R_h] $. Such a polynomial is based on known conditions $M_\infty (R_l) = M_l$, $M_\infty (R_h) =M_h$, $M\,'_\infty(R_l) = 0$, $M\,'_\infty (R_h) =0$. Let us also set $t = (R-R_l) /\Delta R $, where $\Delta R = R_h - R_l$ and $\Delta M = M_l-M_h$. Then the polynomial $S_3 (R)$ becomes 
\begin {displaymath}
    S_3 (R) = \Delta M \left (2 t^3 -3 t^2 \right) + M_l, \quad t \in [0,1]\,.
\end {displaymath}

For further purposes it is convenient to rewrite the equation for local coordinates $t$ corresponding to some given mass $M_\infty$ in the form
\begin {equation} \label {poly}
    2t^3 - 3t^2 + \mu = 0\,,
\end {equation}
where the reduced mass $\mu $ is evaluated under the formula $\mu = (M_l-M_\infty)/\Delta M$. The form and values of the roots of this equation depend on the magnitude of $\mu $, {\it i.e.}, from the relation between three masses $M_\infty $, $M_l$ and $M_h$. It is not difficult to verify that when $0 \leq \mu \leq 1$ the equation \eqref{poly} has three real roots placed on the interval $ [-1/2, 3/2]$.
\myfigures{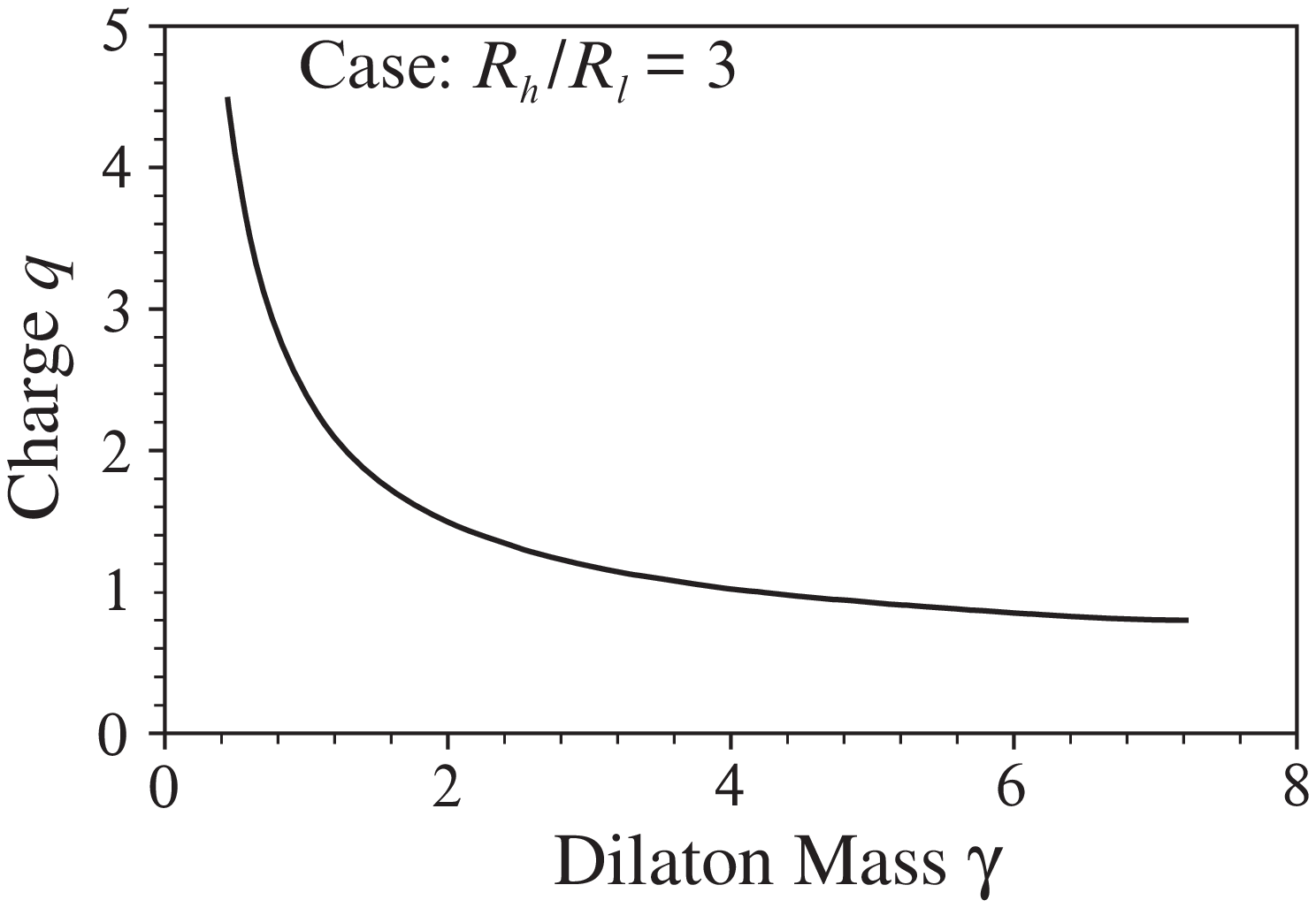}{0.45}{Curve $\rho (q, \gamma) = 3$.}{0.46}{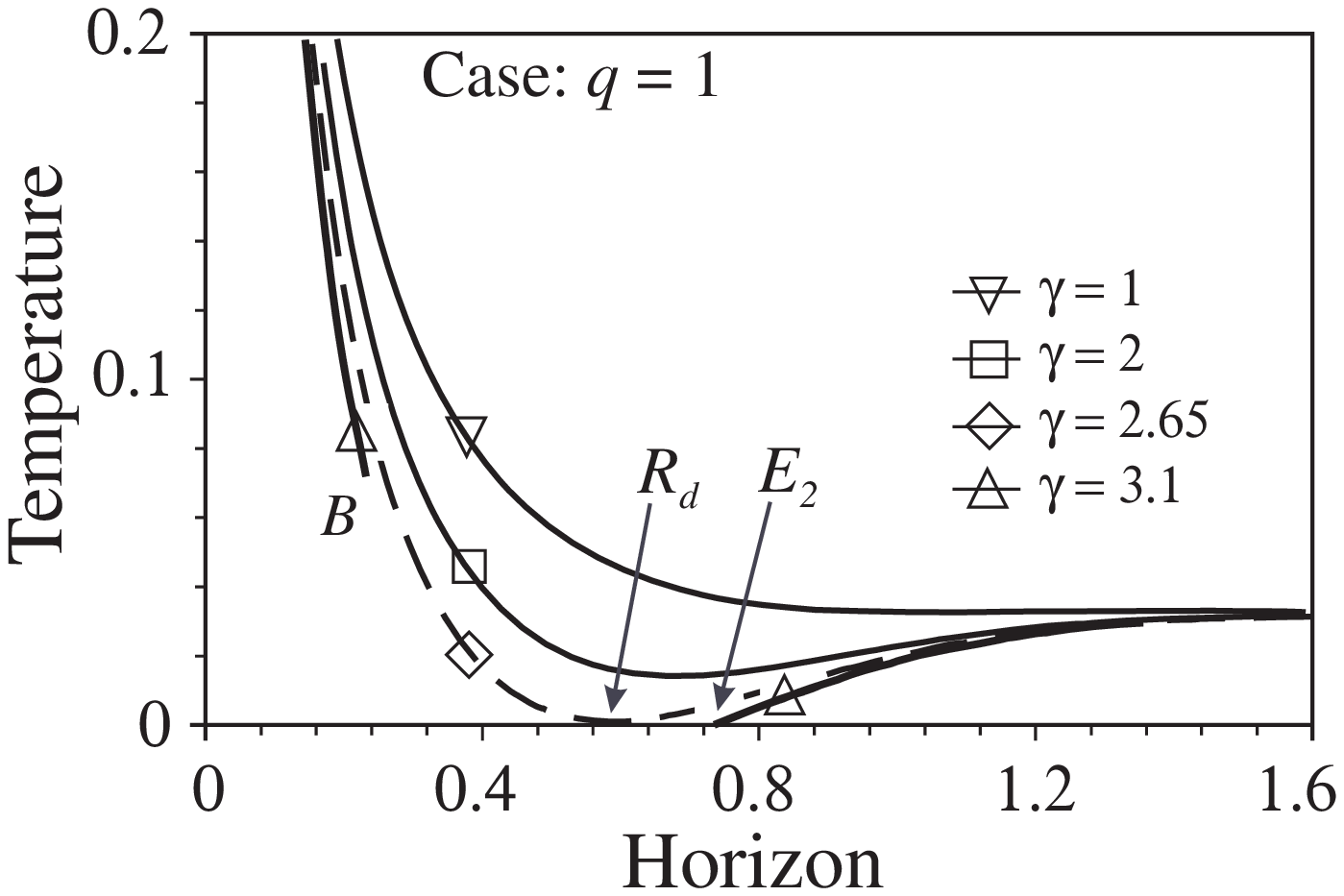}{0.45}{BH temperature.}{0.4}

As it is seen from Fig.\ \ref{scurve}, for large enough $\gamma$ BH can have a regular horizon, only if mass $M_0 < M_h$, which is equivalent to $\mu < 1$. If $\mu > 1$, then according to the above-stated remarks BH has two or three horizons. We continue formally polynomial $S_3(R)$ on the segment $R\in [0, R_l]$ and require mass $M_0$ corresponding to local coordinate $t_0 = -R_l/ \Delta R$, to coincide with extremal mass $M_h$. In this case coefficient $\mu = 1$ and Eq.\ \eqref{poly} possesses one simple root $t_0 = -1/2$, and also two repeated roots $t = 1$. Further we introduce into consideration quotient $\rho \equiv R_h/R_l \geq 1$. Value $\rho = 1$ corresponds to the case of BH with triply degenerated horizon, {\it i.e.}, to the point $D$ in Fig.\ \ref{fig9}. For $\gamma > \gamma_d$ the inequality $\rho > 1$ holds. In particular, for $t_0$ the magnitude $\rho$ possesses the value
\begin {equation*}
    \rho (q, \gamma) = \frac {R_h (q, \gamma)} {R_l (q, \gamma)} = 3\,.
\end {equation*}

The respective dependence between charge $q $ and dilaton mass $\gamma$ is illustrated in Fig.\ \ref{fig13}.

In this way if $1 \leq \rho < 3$, then to the left of extremal mass $M_h$ there is a single-valued branch of dependence $R_h (M_\infty)$. In interval $M_h < M_0 < M_l$ (point $M_0$ is located to the right of point $M_h$ and $\rho > 3$) parameter $\mu \in (0,1)$. From here it follows $t \in \left(-1/2,0 \right)$ and, hence, BH has even two regular horizons.

For calculus of coordinate $R_c $ of point $C $ we continue polynomial $S_3 (R)$ on the segment $R \in [R_h, R_c]$, {\it i.e.}, $t \in [1, (R_c - R_l) /\Delta R \>]$. Then $M_c=M_l$, $\mu=0$ and Eq.\ \eqref{poly} has single nonzero root $t_c = 3/2$, to which corresponds horizon
\begin {equation} \label {rc}
    R_c = \frac {3R_h - R_l}{2}\,.
\end {equation}

The appearance of a triply degenerated horizon with consequent arising of extremal horizons under influence of dilaton mass $\gamma$ one can see in the graph of dependence between BH temperature\footnote {\,Function $\delta (r)$ is a solution of the Cauchy problem \eqref{delta}}
\begin {displaymath}
    T (R_h) = \frac {1} {4\pi} \exp \{-\delta_h \} \, f \, '_h \,
\end {displaymath}
\noindent and external horizon $R_h$ when charge $q $ is fixed (See Fig.\ \ref{fig14}). If BH mass $\gamma < \gamma_d$, the graphs of temperature $T (R_h)$ are smooth curves. For $\gamma = \gamma_d$ the graph $T (R_h)$ concerns the abscissas in point $R_d$, and the BH temperature is equal to zero. If BH mass $\gamma > \gamma_d$, then the temperature curve is splitted to two branches corresponding to arcs $\widehat {AB} $ (left branch) and $\widehat {E_2C} $ (right Branch) in Fig.\ \ref{scurve}.

\section {Conclusion}

In this paper we show, that depending on the form and the number of horizons for equations of symmetrical charged BH in the frameworks of the string Einstein-Born-Infeld model with massive dilaton it is necessary to state different multipoint BVPs with unknown boundaries. For their solution we propose an effective iteration method based on CANM in combination to the method of collocation for treating the arising linearized problems.

The horizons for the extremal BH solutions are found from non-linear algebraic system depending on charge $q $ and dilaton mass $\gamma$. On the graph of horizons against BH mass the presence of extremal horizons is exhibited in appearance of separate branches corresponding to BH with one, two and three horizons. The relevant special case is the single solution of this system, representing the triply degenerated horizon. After solving BVP for BH with extremal horizons and calculation of respective masses of BH the Hermite cubic polynomial is built. The real roots of this polynomial specify the number and form of horizons in general case.

The relation of the above described BH solutions with the physical reality remains an open problem.\\

After the Russian version of this talk was published in \cite{jinr02_1}, 
an e-print by T. Tamaki was published -- \cite{tam_02}. It overlaps  
some part of our work.

\subsection{Acknowledgments}
The authors express their gratitude to Dr. S.S. Yazadjiev and Dr. M.D. Todorov for useful discussions. We are very sorry for the unexpected decision of Dr. S.S. Yazadjiev to interrupt the collaboration with us and with 
the Joint Group of Gravity and Astrophysics\footnote{http:
//webgate.bg/jgga/index.htm} after the publication of the preliminary results in \cite{jinr01_221}.

This work is partially supported by RFBR grant 0001-00617 and Sofia University grants 404/2001 and 3305/2003.


\end{document}